\documentclass[preprint2]{aastex63}
\usepackage{graphicx}
\def \lc{\>\> ,}

\newcommand{\degree}{\mbox{$^{\circ}$}}
\newcommand{\am}{\mbox{\arcmin}}
\newcommand{\as}{\mbox{\arcsec}}

\newcommand{\kms}{\mbox{\,km\,s$^{-1}$}}
\newcommand{\kkms}{\mbox{\,K\,km\,s$^{-1}$}}
\newcommand\cmv{\mbox{cm$^{-3}$}}
\newcommand\cmc{\mbox{cm$^{-2}$}}

\newcommand{\um}{$\mu$m}

\newcommand{\msun}{\mbox{M$_\odot$}}

\newcommand{\co}{$^{12}$CO}
\newcommand{\coo}{$^{13}$CO}
\newcommand{\cooo}{C$^{18}$O}

\def \18OI{[$^{18}$O\,{\sc i}]}
\def \17OI{[$^{18}$O\,{\sc i}]}

\def \HII{H\,{\sc ii}}
\def     \CII{[C\,{\sc ii}]}
\def \13CII{[$^{13}$C\,{\sc ii}]}
\def \NII{[N\,{\sc ii}]}

\def \OI{[O\,{\sc i}]}

\def \Msun{M$_{\odot}$}

\def \lc{\>\> ,}
\def \beq{\begin{equation}}
\def \eeq{\end{equation}}
\begin{document}

\title{Structure of the W3A Low Density Foreground Region}
\shorttitle{Structure of the W3A Foreground Region}
\author{Paul. F. Goldsmith}
\affiliation {Jet Propulsion Laboratory, California Institute of Technology, 4800 Oak Grove Drive, Pasadena, CA 91109, USA}
\author{William D. Langer}
\affiliation {Jet Propulsion Laboratory, California Institute of Technology, 4800 Oak Grove Drive, Pasadena, CA 91109, USA}
\author{Youngmin Seo}
\affiliation {Jet Propulsion Laboratory, California Institute of Technology, 4800 Oak Grove Drive, Pasadena, CA 91109, USA}
\author{Jorge Pineda}
\affiliation {Jet Propulsion Laboratory, California Institute of Technology, 4800 Oak Grove Drive, Pasadena, CA 91109, USA}
\author{J\"urgen Stutzki}
\affiliation {I. Physikalisches Institut der Universit$\ddot{\rm a}$t zu K$\ddot{\rm o}$ln, Z$\ddot{\rm u}$lpicher Strasse 77, 50937 K$\ddot{\rm o}$ln, Germany}
\author{Christian Guevara}
\affiliation {I. Physikalisches Institut der Universit$\ddot{\rm a}$t zu K$\ddot{\rm o}$ln, Z$\ddot{\rm u}$lpicher Strasse 77, 50937 K$\ddot{\rm o}$ln, Germany}
\author{Rebeca Aladro}
\affiliation{Max-Planck-Institut f\"{u}r Radioastronomie, Auf dem H\"{u}gel 69, D-53121 Bonn, Germany}
\author{Matthias Justen}
\affiliation {I. Physikalisches Institut der Universit$\ddot{\rm a}$t zu K$\ddot{\rm o}$ln, Z$\ddot{\rm u}$lpicher Strasse 77, 50937 K$\ddot{\rm o}$ln, Germany}


\begin{abstract}
We present analysis of \OI\ 63 \um\ and CO $J$ = 5--4 and 8--7 multi--position data in the  W3A region and use it to develop a model for the extended low--density foreground gas that produces absorption features in the \OI\ and $J$ = 5--4 CO lines.  
We employ the extinction to the exciting stars of the background \HII\ region to constrain the total column density of the foreground gas.  
We have used the Meudon PDR code to model the physical conditions and chemistry in the region employing a two--component model with high density layer near the \HII\ region responsible for the fine structure line emission, and an extended low density foreground layer.
The best-fitting total proton density, constrained largely by the CO lines, is   $n$(H) = 250 \cmv\ in the foreground gas, and 5$\times$10$^5$ \cmv\ in the material near the \HII\ region.
The absorption is distributed over the region mapped in W3A, and is not restricted to the foreground of either the embedded exciting stars of the \HII\ region or the protostar W3 IRS5.  
The low--density material associated with regions of massive star formation, based on an earlier study by \citet{Goldsmith21}, is quite common, and we now see that it is extended over a significant portion of W3A.
It thus should be included in modeling of fine structure line emission, including interpreting low--velocity resolution observations made with incoherent spectrometer systems, in order to use these lines as accurate tracers of massive star formation.

 \end{abstract}

\setcounter{footnote}{0}

\section{Introduction}
\label{sec:Introduction}
\noindent

\HII\ regions are ubiquitous throughout the galaxy and are signposts of sites of massive star formation.  
Studying their formation and their impact on the interstellar medium (ISM) is important for understanding Galactic star formation and the lifecycle of the ISM.  
Massive young stars are very important sources of UV radiation and stellar winds that have a high impact on their surroundings, shaping and dissipating the environments in which they were born.  
The presence of bright sources of Far- and Extreme-UV guarantees that the surrounding gas will be ionized.
With sufficient absorption, this will transition to neutral atomic and, with further absorption, molecular gas.  
The details of this transition will depend on the strength of these fields, the location of the \HII\ region ({\it e.g.} whether interior or at the edge of clouds), and the density and extent of the associated gas.
For these and other reasons, \HII\ regions have been extensively observed over a wide range of  wavelengths from cm radio emission to UV.  
Among the important probes of this gas in the vicinity of an \HII\ source are the mm, submm, and far-infrared spectral lines of atomic, ionic, and molecular tracers of the physical and dynamical conditions. 

In this paper we focus on understanding the structure of the gas impacted by the W3A \HII\ region in the W3 complex, a site of active star formation, using a variety of tracers that cover a range of physical conditions, density, temperature, abundance, and ionizing radiation. 
We combine observations of \OI, \CII, \NII, and CO $J$ = 5--4 and $J$ = 8--7 spectra, which arise under different physical conditions and thus probe different regions along the line of sight, with model runs of the chemical profiles and spectral line emission to interpret the observations.  

The observations were made using the upGREAT instrument\footnote{upGREAT is a development by the MPI f\"ur Radioastronmie and KOSMA/Universit\"at zu K\"oln,
in cooperation with the MPI f\"ur Sonnensystemforschung and the DLR Institut f\"ur Optische Systeme.} on SOFIA and the chemical and spectral line models have been produced using the Meudon code\footnote{PDR models published in this paper have been produced with the Meudon PDR code (\citep{Lepetit06}, http://ism.obspm.fr).}. 
Almost all of the data used in this paper were presented in a prior paper \citep{Goldsmith21} that focused on modeling the line of sight of the strongest \OI\ emission from a scan through W3A. 
In that paper the emphasis was on modeling just the emission and absorption seen in \OI\ using the Meudon code to model a constant density cloud.  This paper extends that analysis to provide a deeper understanding of the foreground material.

The \OI\ and high $J$ CO emission lines are formed in hot dense regions of the cloud closest to the \HII\ heating source.  
For clouds oriented such that the heating source is on the far side, as seen from the observer, the cooler and less dense gas on the front side can absorb the warmer far side emission.  
In these cases it is possible to determine conditions in the warmer gas producing emission and the colder foreground gas from the line shape.  
In this paper we utilize all the different gas tracers taken at eight positions passing along a straight line through the continuum peak. 
 We also expand our modeling to include a density profile that more realistically represents the cloud transitioning from a dense hot region near the \HII\ source to an extended low density cooler region representing the outer edge of the cloud nearest to the observer. 

While there is no unique model of the cloud toward W3A, the observations are best fit by a hot dense narrow photodissociation region (PDR) with total density\footnote{We use the expression $n$(H) to denote the total density of hydrogen {\bf nuclei,} given by $n$(H) =  $n$(H$^0$) + 2$n$(H$_2)$, where H$^0$ and H$_2$ denote atomic and molecular hydrogen, respectively.}  $n$(H) 
of order 5$\times$10$^5$ \cmv, transitioning over a couple of magnitudes of visual extinction to a low density, $\sim$ 250 \cmv\ cold region about 10 magnitudes thick.  
The strip map reveals strong \OI\ and CO emission near the strongest radio continuum emission from the \HII\ region,  diminishing considerably at the limits of the observed strip.

This paper is organized as follows.  
In Section~\ref{sec:Obs} we review the data used in this paper and present results that bear on the structure of the foreground material in W3.
In Section~\ref{sec:W3} we discuss the W3 region and the constraints on modeling the Photon Dominated Region (PDR) emission and the  absorption produced by the foreground cloud.
In Section~\ref{sec:PDR} we present a PDR model that does a reasonable, albeit imperfect, job of reproducing the spectra seen in emission and absorption. 
We analyze the C$^+$ emission in Section~\ref{sec:CII} and the velocity structure of each tracer in Section~\ref{sec:Velocity}.  
In Section~\ref{sec:Discussion} we draw some general conclusions from the comparison of model and line spectra, and, finally, we summarize our results in Section~\ref{sec:Summary}.


\section{Observational Data}
\label{sec:Obs}
The data used here consist of observations of \OI\ 63 \um, \NII\ 205 \um, CO $J$ = 5-4, and CO $J$ = 8-7 that were presented in \citet{Goldsmith21}, and the reader is referred to that paper for details on the observational procedures and  basic data reduction.  
In that paper, the main focus was on understanding the conditions leading to the very deep absorption observed in \OI\ at the positions of maximum \OI\ emission intensity.  
In the present paper we focus on understanding the properties of all the gas components associated with the main \HII\ region in W3 and seven lines of sight along a strip through the central peak.   
Some additional processing and line profile fitting has been carried out on the \OI\ and CO lines at positions away from the peak  emission, as described below.
We also analyze \CII\ 158 $\mu$m data from \citet{Gerin15} for W3 IRS5, and use this and other \CII\ and radio recombination line data from several studies to develop a more complete picture of the region.


\section{The W3 Region}
\label{sec:W3}
 W3 is a cloud with active star formation and several bright \HII\ regions.  
 It was one of the earliest such regions studied with a variety of tracers of gas properties and \HII\ sources. 
 The structure of a cloud in the presence of bright \HII\ regions consists of an ionized region, a diffuse atomic and possibly molecular layer, a dense PDR, and finally a dense shielded region. 
 As discussed in \cite{Goldsmith21}, we envision that the line of sight to an \HII\ source consists of a PDR powered by UV from a cluster of massive young stars, which also produce extensive ionized gas. 
 An important consideration in developing a model for the region is the size of emission from ionized gas (traced in \NII) and strong PDR emission (traced by \OI).   
 Here we combine observations and cloud models to understand the structure of these layers.  
 In this section we begin with a review of what was previously known about the W3 region and then introduce what we learn from the observations analyzed here.
 
 \subsection{Distance to W3}
 
The distance to W3 has been investigated in many studies employing a variety of techniques. 
\citet{Megeath08} give a useful summary in \S2 of star formation in the region. 
Early studies based on its velocity and using a Galactic kinematic model gave a distance 3.1 kpc  \citep{Reifenstein70} and 2.3 kpc \citep{Georgelin76}.  
The larger distance was used in many studies, but has been replaced with more precise values based on kinematic modeling of H$_2$O masers in outflows \citep{Imai00}, VLBI of maser sources \citep{Xu06, Hachisuka06}, and stellar spectrophotometry \citep{Navarete11}.  
These approaches all give results between 1.83 kpc and 2.0 kpc, and we adopt the latter value in this paper. 
Results from early papers employing larger distances (e.g., \citet{Wynnwilliams71, Beetz74, Beetz76} have been updated using the 2.0 kpc distance, where appropriate.

\subsection{Stars, Ionization, and Extinction}
\label{sec:stars}

The optical nebula IC1795 includes the W3A--D \HII\ regions.
\citet{Wynnwilliams72} observed the W3 region at infrared wavelengths between 1.65 \um\ and 20 \um.
In the strongest and largest well--defined radio continuum source, W3A, they found a compact infrared source at 2.2 \um, which they denoted W3 IRS2. 
They were able to determine its absolute magnitude at 2.2 \um\ to be -5.0 and the foreground visual extinction to this source to be 14 mag.  
They point out that if W3 IRS2 were an O5 star with this foreground extinction, its apparent visual magnitude would be 20$\pm$3 mag, thus explaining why it would not be detected on the Palomar Sky Survey prints.

A second early--type star very close to W3 IRS2 was identified by \citet{Beetz74}, and denoted W3 IRS2a.  
These two potential exciting sources were studied in detail by \citet{Beetz76}.
They confirmed a very large extinction in front of W3 IRS2, $A_v = 15.1$ mag, and also that there is a very large N--S extinction gradient, dropping to as low as $\simeq$ 4 mag.  The extinctions derived assume a distance of 3.2 kpc. 
If the more recently derived distance of 2 kpc is used with the observed flux densities and derived O5 V spectral type for IRS2, the attenuation by the foreground material must be a factor $(3.1/2.0)^2$ larger, corresponding to an increase of about 1 mag. in the extinction. 

The column density of H$_2$ and dust in W3 does not peak at the position of these massive, embedded stars, but rather at the position of the embedded protostar W3 IRS5.  
This association is seen very clearly in Figure 8 of \citet{Rivera13}, which is the basis for Figure 3 in \citet{Goldsmith21}.  
The offset between IRS2 and IRS5 is clear in the 30 \um, 50 \um, and 100 \um\ data of \citet{Werner80}, and dramatically evident in the distribution of total far--infrared luminosity these authors derive.  
W3 IRS5 has negligible radio continuum emission compared to that from W3A.  
It has been the subject of extensive infrared and molecular observations, revealing it to be a deeply embedded, young OB protocluster \citep{Tieftrunk95,Megeath96,vanderTak05}.

\subsection{\HII\ Regions and Ionized Gas in W3}
\label{sec:ionized}
   
To study the ionized gas in W3 \citet{Wynnwilliams71} observed the free--free emission at 4995 MHz using the Cambridge One--Mile interferometer which yielded a FWHM beam width of 6.5\as.  
This beam size was sufficient to resolve the strongest individual concentration, W3A, its closest neighbor, W3B, and a somewhat more distant source, W3C.  
An additional concentration, W3D, was also observed, along with weaker, more diffuse emission.  
To determine the total column density \citet{Wynnwilliams71} derived the optical extinction in front of the ionized gas by  studying the stars that they identified as the exciting stars of the \HII\ regions.  
They found that the visual extinction was highly variable, from $\simeq$ 3 mag. to almost 15 mag.  
We will use this information when comparing extinctions to CO column densities, and when constructing models of the foreground gas.

The W3 region was later observed by \citet{Colley80}  at 2.7 GHz and 15.4 GHz, with a maximum angular resolution of 0.65\as.  
They confirmed numerous \HII\ regions found in earlier studies, and developed a model in which the massive stars responsible for the ionized gas formed in a neutral gas cloud. Their model shows that the ionized gas has a shell--like structure due in part to the confinement by the neutral gas remaining outside the \HII\ region.

Subsequent radio continuum studies of W3 extended the frequency range covered and pinpointed the locations of the infrared sources relative to the ionized gas. 
\cite{Salter89} observed at 90 GHz and 140 GHz, while \cite{Tieftrunk97} observed at 5, 15, and 22.5 GHz. 
 The very high sensitivity and angular resolution (0.1\as)  afforded by the VLA used by \citet{Tieftrunk97} allowed imaging of the H $66\alpha$ radio recombination line as well as the radio continuum emission.  
From Figure 9 of that paper, we see that the IR Source W3 IRS2 is close to the center of the emission from the ionized gas.  

Defining the size of an \HII\ region is challenging due to the absence of a sharp edge for the radio continuum emission, the varying electron densities derived at different frequencies, and the distance uncertainty.  
All of these add to the uncertainty in the mass.  
The  \HII\ region W3A  has angular dimensions \citep[][Fig. 6]{Tieftrunk97} of approximately 50\as\ x 90\as, corresponding to an average linear dimension of 0.75 pc at a distance of 2 kpc.
From observations at frequencies of 5, 15, and 22.5 GHz, scaled to distance of 2 kpc using the expressions of \citet{Panagia78}, \citet{Tieftrunk97} derive a mass of 20.0 \Msun, with only modest differences at the various frequencies observed.
The average electron density is 5.4$\times$10$^3$ \cmv, with the peak electron density approximately 40\% greater.

The electron density and emission measure are consistent with those found previously by \citet{Colley80}. 
One significant difference is that Figure 5 in \citet{Colley80} shows very clear, low--level emission extending to the Northeast of W3B.  
This feature is not seen in Figure 9 of \cite{Tieftrunk97} or images in several other studies.
This suggests that some low-level extended emission may have been removed by the interferometric observations and data reduction.  
This emission is important because it underlies the most extreme northwest position towards which we detected \NII, which is reasonable given that strong ionized nitrogen  emission is expected to be produced only in fully ionized regions.

\begin{figure*}[htb!]
\includegraphics[angle=0, width=16 cm]{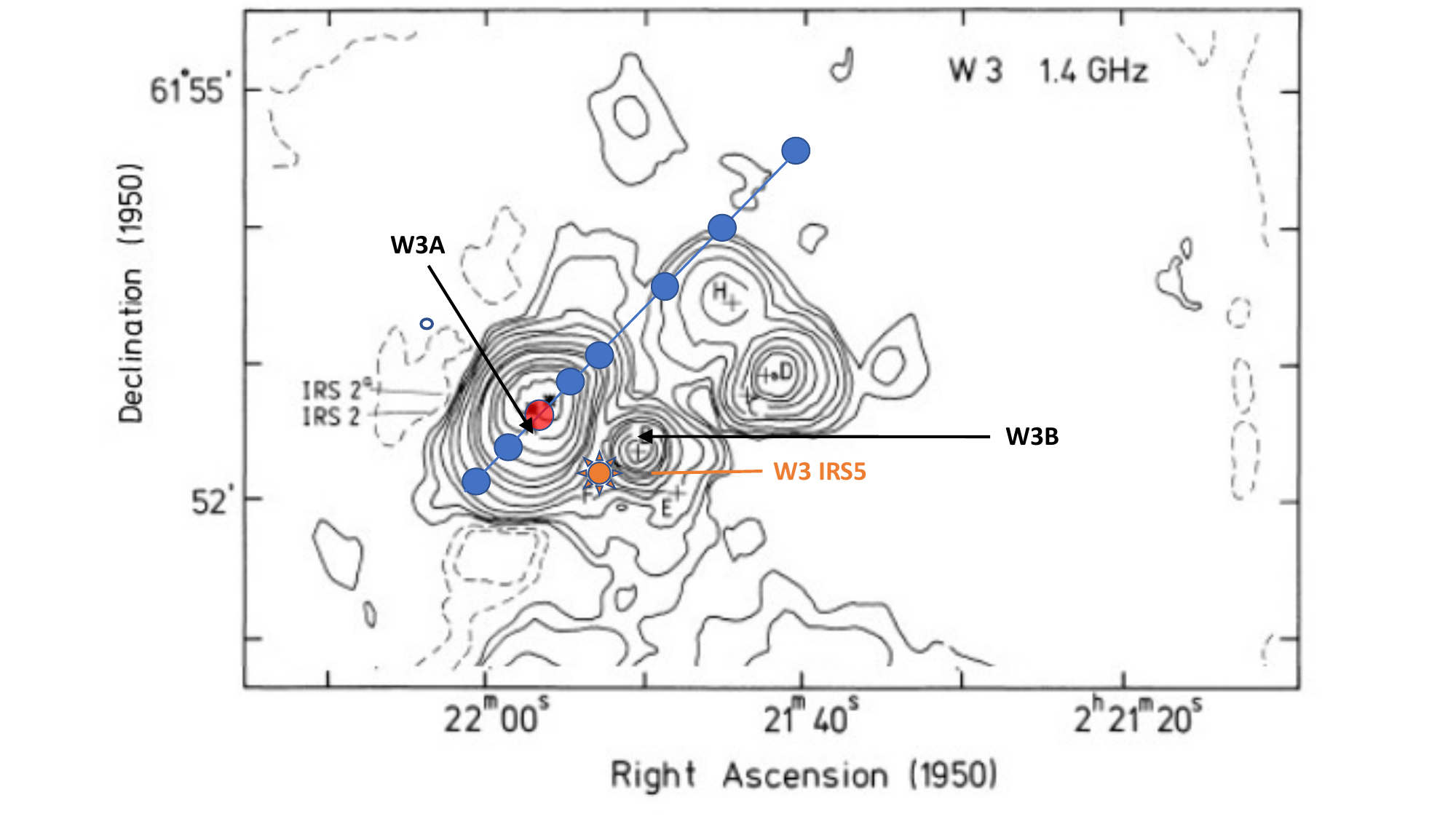}
\centering
\caption{ Free--free emission at 1.4 GHz from the W3 Main \HII\ region. This figure, adapted from Figure 3 of  \cite{Roelfsema91}, includes several strong emission peaks.  There is extended, weaker emission to the south which is not shown here.  The two exciting sources IRS2 and IRS2$^a$ are shown; they are both essentially coincident with the peak emission of the continuum source denoted W3A.  The central position of our strip is nearly coincident with these two IR sources which collectively power the \HII\ region.  The eight positions observed in \NII\ 205 \um, \OI\ 63 \um\ and CO $J$ = 8-7 and $J$ = 5-4 are indicated by the filled blue circles except for the filled red circle denoting the central peak position).  The highly--embedded infrared source W3 IRS5 is indicated by the filled orange circle.\label{fig:1.4GHz}
}
\end{figure*}

In this paper we use the emission from the fine structure \NII\ line at 205 \um\ to trace the ionized gas.  Nitrogen has an ionization potential of 14.53 eV, so it traces the extreme ultraviolet (EUV) environment associated with \HII\ regions.  
Figure \ref{fig:1.4GHz} shows the map at 1.4 GHz and is adapted from \cite{Roelfsema91}.  Marked on the figure are the 8 positions observed in \NII\  205\um. 
The emission line data were presented in Figure 5 and Table 3 of \citet{Goldsmith21}.  
The integrated \NII\  intensities are strong at five positions in W3A, but decrease significantly at the two positions to the Northeast at the edge of and beyond the source W3H.  

The \NII\  intensities are not highly correlated with the strength of the free--free emission.  
Rather, the \NII\ appears to originate from less--structured ionized gas that presumably includes the condensations W3A and W3H.  
The extent of the strong \NII\ emission is, however, well-defined by that of the radio continuum emission, confirming that both of these trace the fully-ionized gas in W3A.
This distribution is consistent with the 2.7 GHz map of \citet{Colley80}, that has significant extended emission to the west and north of W3A.

While singly--ionized nitrogen is expected to be the dominant form of this element in \HII\ regions, the early stellar types of IRS2 and IRS2a make it possible for these sources to produce significant doubly--ionized nitrogen in their vicinity.  
This issue is discussed in $\S$5.5 of \citet{Goldsmith21}, where the column density of N$^{2+}$ increases the total nitrogen column density by a factor $\simeq$1.6, yielding good agreement with that expected from the hydrogen column density.
Overall, the $\simeq$ 2 pc extent of the \HII\ emission is consistent with that of the radio continuum emission from W3A, allowing for a low--level, extended component together with the clear peaks seen in high angular resolution studies.

There are significant changes in the \NII\ line width and velocity centroid at different positions  \cite[see Figure 5 and Table 3][]{Goldsmith21}. 
It is difficult to correlate the changes in these line parameters with any other aspects of the structure of the ionized gas; the issue of the velocities relative to that of other components is discussed {\bf below} in $\S$ \ref{sec:Velocity}.


\subsection{Extended Molecular and Dust Distribution}
\label{sec:Largescale}

The large--scale structure of the W3 region has been the subject of numerous studies based on emission from dust and low--lying rotational transitions of carbon monoxide.
 \cite{Lada78} conducted a large--scale survey of \co\ $J$ = 1-0 covering 16 square degrees with 8\arcmin\ resolution and derived a total CO traced mass of $\simeq$10$^5$ \msun.  
Additional observations of $^{13}$CO with 2.6\arcmin\ resolution helped determine the mass of the major condensations.
These density enhancements, including W3A, fall approximately on a line in Galactic latitude (not far from a line in Right Ascension) on the Western edge of ionized gas of the W4 \HII\ region. 
The local maxima are presumed to be a filament or sheet compressed by the star formation in W4, and denoted the ``High Density Layer'' (HDL). 
The connection between W4 and W3 is also made clear by the larger--scale CO study of \citet{Heyer98}.

A more limited region was observed in \co, \coo, HCN, and CS by \cite{dickel80} at 70\as\ to 150\as\ resolution and position--dependent sampling interval.  
\cite{dickel83} combined these molecular data  with radio and optical data to develop a model for the region.
Their model for W3A was a very young  blister \HII\ region  breaking out as a ``Champagne Flow.''

A  $\simeq$1 square degree region was mapped in \co\ and \coo\ $J$ = 2--1, together with $J$ = 3--2 \co\ observations of just W3 Main by \cite{bieging11}.  
These data confirm the earlier observations by \citet{Brackmann80} which revealed that the low--$J$ CO transitions show self--absorption. 
The peak of the \co\ and \coo\ $J$ = 3--2 and $J$ = 2--1 emission is at the position of the highly--embedded protocluster, W3 IRS 5, separated by about 30\as\ (0.3 pc) from the central position of our strip.  
The self--absorption extends over a large portion of the central region of W3, indicating that whatever gas is responsible for the absorption is also highly extended, rather than being concentrated in front of W3 IRS 2 or IRS 5.  

A region $\simeq$1\degree\ x 1\degree\ in size was mapped by \cite{polychroni12}, but including the $J$ = 3--2 transitions of \co, \coo, and \cooo.  
The detectable \co\ emission extends over a region $\simeq\ $0.8\degree\ x  0.8\degree\ on the sky, with the strongest CO $J$=3--2 emission (their Figure 2) being closely restricted to the Eastern edge as is the $J$ = 1--0 emission observed by \cite{Lada78}.  
The multiple transitions, including those not significantly affected by optical depth, allowed a more accurate determination of the gas mass of this region to be 4.4$\pm$0.4$\times$10$^5$\msun.  
The mass distribution of this extended cloud is nonuniform, but on a large scale is concentrated at the Eastern HDL boundary, with more modest column density enhancements along almost the entire periphery of the region mapped.
From the above mass and assuming a line of sight dimension equal to the region's transverse extent (28 pc), the average column density in the W3 region would be <$N$(H$_2$)> $\simeq$ 2$\times$10$^{22}$ \cmc.

To determine the location of the \HII\ region within the cloud, it is valuable to get an idea of the column density distribution in the central region of the cloud.
The peak H$_2$ column density in the central region of W3 (including W3 IRS5) found by \cite{Rivera15} in a detailed study  based on multi--wavelength dust emission mapped by{\it Herschel} is approximately 2$\times$10$^{23}$ \cmc.  
The column density at the positions in our strip with the strongest \OI\ emission is about half this value, and thus dramatically higher than that of the W3 cloud as a whole.
The average density in the W3 region would be $< n({\rm H}_2)>$ $\simeq$ 200 \cmv, compared to $<n({\rm H}_2)>$ = 5.6$\times$10$^4$ \cmv\ for the central peak region (nominal depth of 1.2 pc based on averaged line of sight dimensions) mapped by \cite{Rivera15}.

The column density, $N$(H$_2$) = 1.5$\times$10$^{22}$ \cmc\ in front of the \HII\ region (implied by the extinction to the embedded sources discussed in \S3.2) is thus far below its peak value that occurs at a position separated by only $\simeq$0.8 pc on the sky.
The dust--determined column density measured by far--infrared emission traces the entire line of sight column, while that to the W3A \HII\ region measures only that to the stars and the surrounding ionized gas.  
It thus appears that the \HII\ region is located relatively close to the boundary of the W3 cloud facing the Earth, with only about 10\% of the material along the line of sight in front of it.
The values of column density and density determined by the extinction measurements are used for modeling the relatively low-density foreground gas presented in \S \ref{sec:PDR}.

The inhomogeneous structure of the W3 region is very clearly shown in the 850 \um\ map of \citet{Moore07}, who identified 316 dense clumps having masses between 13 and 2500 \Msun.
Within this 0.8 square degree region, the continuum peaks are generally within the region defined by the CO $J$ = 1--0 emission, but the likely presence of major temperature variations make it difficult to see a close correlation even if one were present.
In the central region of W3A, \citet{Rivera15} also found that the morphology of the dust column density and dust temperature are quite different.  
These results suggest that there may well be significant column density variations on different scales along the line of sight throughout W3 as well as from one line of sight to another.\\

\subsection{Low--Excitation Foreground Gas}
\label{sec:Lowex}

\begin{figure*}[htb!]
\includegraphics[angle=0, width=14 cm]{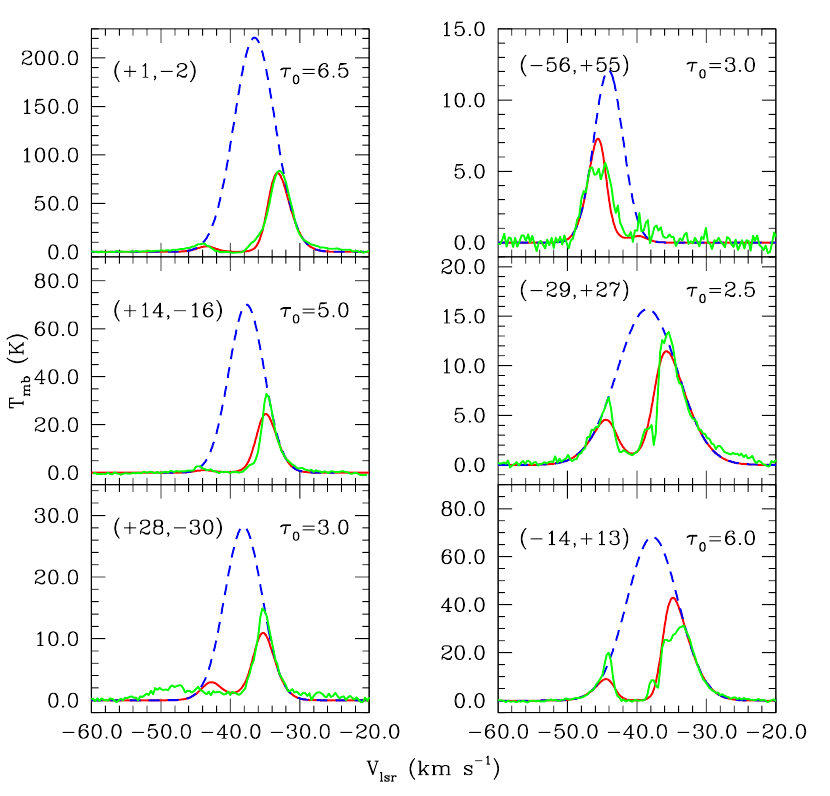}
\centering
\caption{Spectra of \OI\ 63 \um\ at six positions in W3 with strong emission and discernible absorption.  The position of each spectrum is given by the offsets in arcseconds relative to the central position, denoted by  filled red  circle in Figure \ref{fig:1.4GHz}, having J2000 coordinates $\alpha$ = 02$^h$25$^m$ 44.5$^s$, $\delta$ = +62$\degree$06$\am$11.7$\as$. The observed spectra are indicated by the green lines.  The background spectra -- those that would be observed in the absence of foreground absorption -- are obtained by fitting the line wings and shoulders as discussed in the text and in the Appendix, and are shown as the dashed blue curves.  The simulated observed spectra, obtained by adding a single foreground absorption component with Gaussian line profile and optical depth indicated in upper right of each panel to the background spectra, are shown as the red curves.  
\label{fig:OIspec}
}
\end{figure*}

As discussed in \citet{Goldsmith21}, there is clear evidence for absorption in the \OI\ 63 \um\ line.  
While this may be considered to be ``self--absorption,'' it is more correct to call this foreground absorption since the physical conditions in the absorbing gas are very different from those in the gas producing emission.  
Absorption is absent at the most extreme Northwest position ($\Delta \alpha, \Delta \delta)$ = (-113,+112) (the  offsets are in seconds of arc from the central position) and is marginally visible at the adjacent position (-85,+84).
This behavior is in part a reflection of the intensity of the \OI\ emission, in addition to the weakening of the absorption at the edge of the region. 
For  six of the positions shown in Figure \ref{fig:1.4GHz}, there is clear evidence for \OI\ absorption; the \OI\ spectra for these positions are shown in Figure \ref{fig:OIspec}.

Overall, the extent of the strong \OI\ emission, evident in Figure 7 and Table 2 of \citet{Goldsmith21}, is very close to that of the \NII\ emission supporting the picture that the hot portion of the photon dominated region is powered by the ionized gas, which in turn is powered by the embedded stars ($\S$ \ref{sec:ionized}).  
At a distance of 2 kpc, the minimum extent in the plane of the sky of the strong \OI\ 63 \um\ emission is 1.5 pc.  
We obviously do not have complete sampling of this component of W3, but it seems reasonable to model the emission and absorption components as spatially extended, and certainly not restricted to an individual line of sight.

The six positions with clear \OI\ foreground absorption (Figure \ref{fig:OIspec}) have been analyzed in the following manner.   
First, to represent the background emission that would be visible were there no foreground absorption (configuration with the source heated by the \HII\ region on the near--side of the extended, lower density component) we fit a Gaussian  to the observed line wings, adjusting the velocity range to maximize the peak intensity and minimize the differences between the Gaussian and line wings  (the details of this approach are discussed in the Appendix).  
This process works very well, in part due to the inherently Gaussian form of the strong PDR \OI\ emission, together with the fact that the absorption line width is less than that of the emission, allowing the fit to be made over a sufficient velocity range to determine it quite precisely. 

Strictly speaking the solutions are lower limits to the peak intensity of the near-side \OI\ emission.
We list the Gaussian fitting parameters (as defined in the Appendix) inTable~\ref{tab:W3_all_gaussfits}, which is an updated version of Table 3 in \citet{Goldsmith21} and now includes the 1-sigma errors.  For some positions, for example, at offset (+28,-30), there are clearly additional emission features at velocities beyond the single Gaussian representing the strong PDR emission, and we exclude these velocity ranges from the fit.
\begin{deluxetable*}{cccccc}[ht!]			
\tablecaption{W3 \OI\ 63 \um\ Spectral Line Gaussian Fits
\label{tab:W3_all_gaussfits}}															
\tablehead{
\colhead{Offsets (\as,\as)}
&\colhead{ $T_0$ \tablenotemark{a}}
&\colhead{$T_{p}$\tablenotemark{b}}
&\colhead{$V_{p}$\tablenotemark{c}}
&\colhead{$\delta V_{FWHM}\tablenotemark{(b,d)}$}
&\colhead{$\int T_{mb}dv$}\\
\colhead{($\Delta \alpha$, $\Delta \delta$)}
&\colhead{(K)}
&\colhead{(K)}        
&\colhead{(\kms)}
&\colhead{(\kms)}
& \colhead{(\kkms)}
}
\startdata
(-56,+55)& 0.2 &11.7$\pm$1.8&-43.9&5.2$\pm$0.3&60.8\\
(-29,+27)&0.2 &15.7$\pm$0.4&-38.5&10.2$\pm$0.1&160.4\\
(-14,+13)&0.1 &68.3$\pm$1.4&-37.8&8.7$\pm$0.1&596.2\\
(+1,-2$~$)&1.7 &220.9$\pm$10.2&-36.5&7.0$\pm$0.1&1538.2\\
(+14,-16) & 0.6 &62.7$\pm$5.5&-38.8& 7.9$\pm$0.2&497.8\\
(+28,-30)& 0.6 &28.2$\pm$2.4&-38.2&6.3$\pm$0.2&178.6\\
\enddata
\tablenotetext{a} {$T_0$ is a constant offset that is due to residual values after baseline removal from a larger velocity range. It could represent a residual continuum emission.}
\tablenotetext{b} {The uncertainties are the 1--sigma outputs of the Gaussian fitting routine.} 
\tablenotetext{c} {The fitting errors for the central velocity, $V_p$, are all less than 0.13 \kms.}
\tablenotetext{d} {$\delta V_{FWHM}$ is the full width half maximum of the best--fit Gaussian.}
\end{deluxetable*}

Second, we fit another Gaussian function representing the absorption. 
The peak optical depth is a free parameter, and the line width and velocity centroid are also allowed to be different from those determined for the emission profile. 
The radiative transfer is handled assuming that there are two independent slabs, and the emergent signal is compared to the observed data to determine the best fit parameters for the absorbing gas.
We see in Figure \ref{fig:OIspec} that the best--fit model spectra including the fitted emission and absorption components reproduce the observed spectra very well. 
As the predicted intensity of the source of \OI\ on the near--side is a lower limit (see Appendix) the derived opacity for the absorbing layer is strictly a lower limit.
 The only systematic differences between model and data are the presence (mentioned above) of additional emission features, and ``sharper--peaked'' emission features on either side of the main absorption dip.   
 One possible explanation for the latter is that there are multiple absorption components with slightly different velocities.

The values of the best fit representation of the emission line without foreground absorption and the parameters of the  absorption in terms of a single Gaussian are given in Table \ref{tab:fgcol}.  
These are the peak background $T_A$(K), peak foreground, peak foreground optical depth, the line width of the foreground absorbing gas, and the foreground column density N(O$^0$).
The column densities of atomic oxygen in the foreground component are obtained assuming zero excitation, equivalent to assuming that all atoms are in the $^3P_2$ state, which is valid if T$_{ex}$(\OI\ 63 \um) $\ll$ $h\nu/k$ = 227.7 K.  
This assumption is further discussed and justified in $\S$\ref{sec:PDR}. The resulting expression for the  atomic oxygen column density (in cm$^{-2}$ 

\begin{equation}
N({\rm O}^0) = 1.91\times10^{17}\int \tau(v)dv \lc
\end{equation}
which assuming a Gaussian line profile becomes (see Equations 22 and 23 of  \citet{Goldsmith19})
\begin{equation}
N({\rm O}^0) = 2.03 \times10^{17} \tau(v_0) \delta v_{FWHM},\,
\end{equation}
\noindent where line widths are in km s$^{-1}$, $\tau (v_0)$ is the optical depth at the peak velocity ($v_0$) of the Gaussian describing the line profile, and $\delta v _{FWHM}$ is the FWHM of the Gaussian.

\begin{deluxetable*}	{ccccc}[ht!]
													
\tablecaption{Foreground Atomic Oxygen Column Density at Six Positions in W3
\label{tab:fgcol}}
										
\tablehead{
\colhead{Offsets\tablenotemark{a}}                                                                 &\colhead{Peak Background }  
& \colhead{Peak Foreground Optical Depth  \tablenotemark{b}}         &\colhead{Foreground Line Width} 
& \colhead{Foreground $N$(O$^0$)\tablenotemark{c,d}}\\
\colhead{($\Delta$$\alpha$, $\Delta$$\delta$)}               & \colhead{ $T_A$ (K)}                                    
&\colhead{ ($\tau_0$)}                                                                         & \colhead{(FWHM \kms)}                            
& \colhead{(10$^{18}$ $\cmc$)}
}
\startdata
(-56,+55)			&  $~$12.1  	& 3.0$\pm$0.4		& 3.8		&   $~$2.3$\pm$0.3\\
(-29,+27)			&  $~$15.7  	& 2.5$\pm$0.2		& 4.5		&   $~$2.3$\pm$0.2\\
(-14,+13)			&  $~$68.3  	& 6.0$\pm$1.0		& 4.5		&   $~$5.5$\pm$0.9\\
(+$~$1, -$~$2)   	&220.9  		& 6.5$\pm$1.0    	& 5.5  	& 	$~$7.3$\pm$1.1\\
(+14,-16)			&   $~$70.0 	& 5.0$\pm$1.0    	& 5.5  	& 	$~$5.6$\pm$1.1\\
(+28,-30)			&	$~$28.2 	& 3.0$\pm$0.2		& 4.8		&   $~$2.9$\pm$0.2\\
\enddata
\tablenotetext{a}{In seconds of arc relative to central position  $\alpha$(2000) = 02$^h$25$^m$ 44.5$^s$, $\delta$(2000) = +62$\degree$06$\am$11.7$\as$} 
\tablenotetext{b}{Combined uncertainties due to background and foreground temperature uncertainties (see text)}
\tablenotetext{c}{Assuming all O$^0$ is in the ground state, as discussed in the text}
\tablenotetext{d}{Uncertainties propagated from those in the absorption optical depth given in column 3}
\end{deluxetable*}
%
%
Due to the fact that we are fitting absorption of an asymmetric background emission profile by an imperfectly--Gaussian (based on visual inspection) absorption line profile, the fitting could not profitably be carried out using a standard Gaussian fitting profile.  
Rather, the fits were carried out by hand, attempting to find the best fit to channels in a $\simeq$5\kms\ region around the minimum. 
The region width was based on typical ranges over which the observed signal was close to its minimum value.  
Three of the positions show minima consistent with zero signal, while the three others have well--defined residual signal levels  (Figure \ref{fig:OIspec}).  

There are at least two contributions to the uncertainty in the derived optical depth of the foreground gas.  
The first is the uncertainty in the intensity of the background emission.
Assuming we are attenuating the background with no emission, the peak optical depth $\tau_0$ is given approximately by the expression $\tau_0 = ln(T_{bg}/T_{min})$, where $T_{bg}$ is the background antenna temperature and $T_{min}$ is the minimum observed temperature.  
In this situation the change in optical depth produced by a change in the background temperature is given by $d\tau = dT_{bg}/T_{bg}$; thus equal to the fractional change in the background.
Taking the 1$\sigma$ uncertainties and fitted values from column 4 of Table \ref{tab:W3_all_gaussfits}, we find that the resulting changes are small, between 0.02 and 0.15 for the six positions showing clear absorption.

With the same assumption of simple foreground absorption, we find $d\tau = -dT_{min}/T_{min}$; thus equal to the fractional change in the minimum signal.  
For the first two and the last position in Table \ref{tab:W3_all_gaussfits}, the minimum values are between 0.5 K and 1.0 K.  
Taking the uncertainty as the rms noise, the resulting uncertainty in $\tau_0$ is between 0.2 and 0.4.  
These values have been incorporated into column 3 of Table \ref{tab:fgcol}, by combining them in quadrature with the uncertainties due to background temperature.

The three positions with stronger absorption and larger optical depth pose a challenge, since in some cases, such as (+1,-2), the actual value of $T_{min}$ is negative.  
This could result from a combination of baseline fitting uncertainties and the more physical effect that the background emission includes a continuum term from the hot dust in the PDR.  
A continuum is strongly suggested by the observations, and is quite reasonable for 63 \um\ wavelength, but the spectral line receivers used do not have sufficiently high continuum stability to make this measurement definitively. 
Given the instrumental and model issues, we assign an uncertainty in optical depth of 1 to these three positions.
The uncertainties in the optical depth have been propagated to yield the uncertainties in the foreground absorbing column densities included in column 5 of Table \ref{tab:fgcol}.


\section{PDR Model}
\label{sec:PDR}

To understand the structure of W3 we have combined our observations with PDR models of the  \OI\ emission and absorption, together with the emission and absorption features in CO(5--4) and CO(8--7) lines. 
To interpret our results,  we have used the Meudon PDR Code \citep{Lepetit06}, which incorporates a plane--parallel geometry. 
The Meudon PDR code solves for steady--state chemistry and thermal balance as a function of distance into the cloud with input parameters consisting of the density profile, UV radiation field at the boundary, and cosmic ray ionization rate.  
The code does not include ionized material, so that we cannot use it to analyze the \NII\ emission, which arises in the fully--ionized gas.  

We consider the heating source of the PDR region to be on the far side of the cloud, based on conclusions of \citet{Goldsmith21}, and as reinforced by the discussion above regarding foreground absorption of tracers \OI\ and CO(5--4) which arise in hot and warm regions, respectively. 
The PDR model described in \citet{Goldsmith21} was a uniform density model, which was adequate to model the \OI\ emission and absorption because the high density, $n$(H)$\simeq$ 10$^5$ -- 10$^6$ \cmv, hot gas produced strong background \OI\ emission, while a dense colder foreground was sufficient to provide the \OI\ absorption.  
As discussed in our earlier paper on W3, the dramatic drop in temperature at more than a few mag. of visual extinction  from the heating source results in very low fractional population of the upper ($^3P_1$) level of the 63 \um\ atomic oxygen fine structure level despite the high density.  
Thus, once the kinetic temperature is significantly less than  $\Delta E/k$ = 227.7 K, we will see absorption of the 63 \um\ line.  
However, a single high density model cannot explain the CO high $J$ emission and absorption. 
Here we expand on the models presented in \cite{Goldsmith21} by implementing a density profile.

The flexibility of  the Meudon model to incorporate a very large number of adjustable parameters raises a challenge as how to best match models and observations.   
We have not tried to create an extensive library of solutions and numerically determine the ``best fit'' model.  
Rather, guided by intuition and experience in previous PDR modeling of fine structure and molecular lines we first fixed the thickness of the high--density layer and its heating, in order to produce the strong \OI\ emission observed in sources without low--excitation foreground layers, and derived by fitting the line wings and shoulders of most positions in W3 discussed above and shown in Figure \ref{fig:OIspec}.
The thickness of this layer is not critical as the \OI\ emission is optically thick, but is important for reproducing observed \CII\ emission seen in other studies \citep[e.g.,][]{Guevara20}. 
This result can be seen in the almost complete independence of antenna temperature on thickness for the 63 \um\ \OI\ line emitted from a slab without low--excitation foreground gas shown in Figure 13 of \citet{Goldsmith21}.

\subsection{Constraints from Extinction}

To constrain the foreground properties we use extinction measurements.  
As discussed in \S\ref{sec:stars} above, the total extinction to the star (or stars) powering the W3A \HII\ region is $\simeq$16 mag, assuming a distance of 2 kpc.  
From the discussion of the \HII\ region itself in \S\ref{sec:ionized}, the total electron column density through W3A is $\simeq$10$^{22}$ \cmc, and assuming a fully ionized gas, is equal to the column density of protons, and implies an extinction of $\simeq$5 mag. in the ionized gas. 
It would be surprising if the existing stars were very far from the center of the \HII\  region, so the contribution to the extinction from dust in the \HII\ region is $\simeq$ 2.5 mag, leaving 13.5 mag in the PDR and any foreground material.  
For purposes of modeling the foreground gas we adopt a slightly larger value of 15 mag. visual extinction, as this is not critical and allows for the possibility of somewhat smaller extinction from dust in the \HII\ region.

\subsection{Constraints from CO Lines}

As discussed above, the excitation of the \OI\ $^3P_1$ level requires a high density and temperature, such that the \OI\ absorption does not constrain the density and temperature in the foreground but only places upper limits on them. Instead, the CO(5--4) and CO(8--7) lines, which are excited at lower temperatures and much lower densities provide insight on density and temperature in a portion of the foreground.  
The energy of the upper $J $ levels for CO(5--4) and CO(8--7) lie at 82.97K and 199.11K for $J$ = 5 and 8, respectively.  
The corresponding critical H$_2$ densities for de-excitation from these levels are $\sim$7$\times$10$^4$ \cmv\ and $\sim$2$\times$10$^5$ \cmv\ for $J$ = 5 and 8, respectively.   Thus these CO lines arise in warm and dense layers close to the region that gives rise to \OI\ emission, while their absorption occurs in slightly cooler, but still warm, and much lower density regions.  \\

\begin{figure*}[htb!]
\includegraphics[angle=0, width=14 cm]{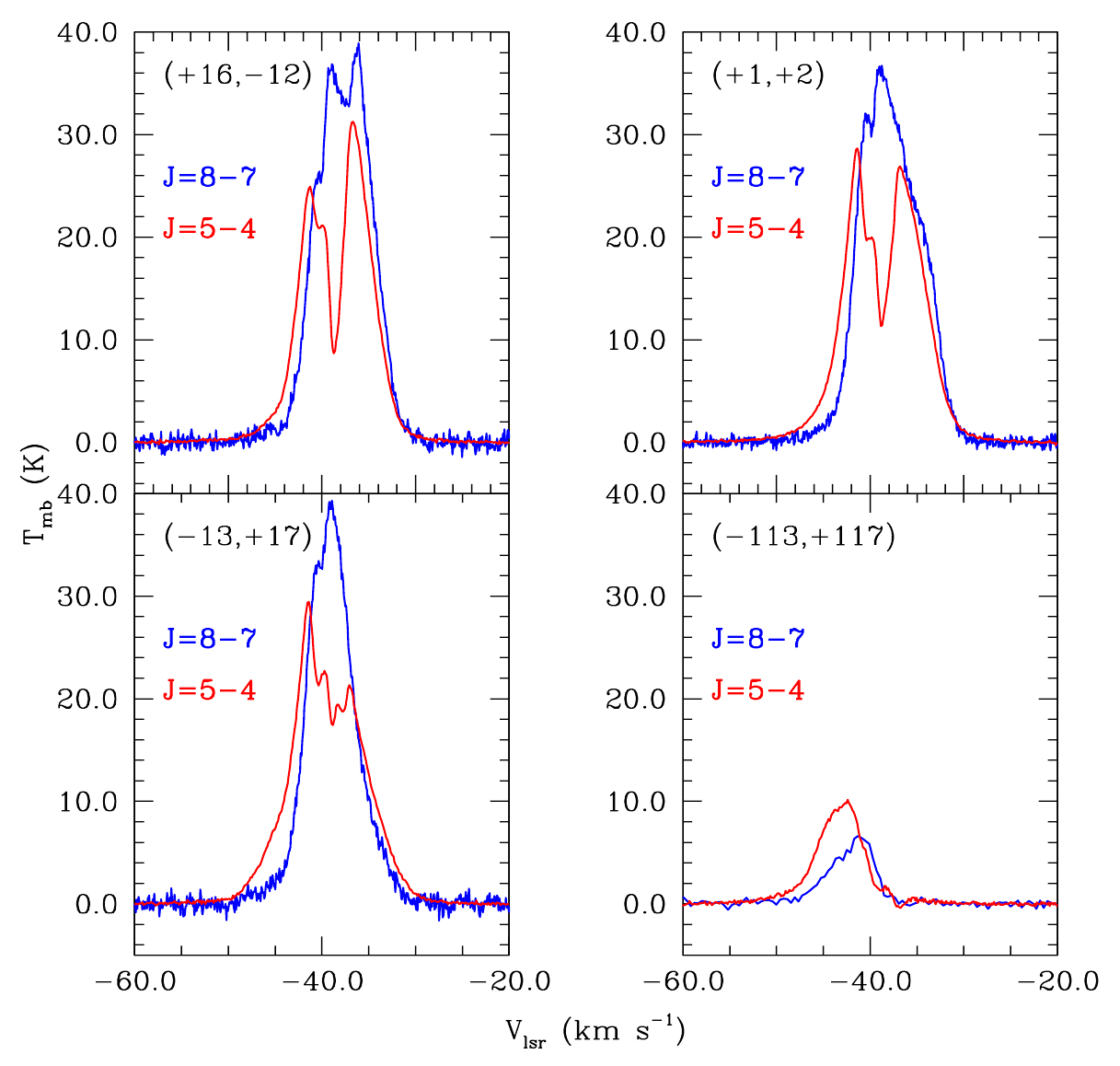}
\centering
\caption{ Spectra of the CO $J$=8--7 and 5--4 lines at four positions in W3.  The three positions near the center of the strip show prominent self--absorption in the $J$=5---4 line, but much less so, or not at all, in the higher $J$=8--7 line.  The fourth position at the extreme end of the strip (-113,+117) does not show self--absorption in either transition.  
}
\label{fig:co4pos} 
\end{figure*}

In Figure~\ref{fig:co4pos} we plot the spectra of the CO(5-4) and CO(8-7) lines at four positions using the offset nomenclature to label them.  
It can be seen that three of the four lines of sight have strong emission and that the CO(5-4) is significantly absorbed, and the CO(8-7) less so.  
Due to the different temperatures and densities required to populate the $J$ = 4 and 7 levels, the absorption of these two CO lines likely occurs in different layers of the foreground gas.

We use our CO observations to model portions of the cloud in front of the \HII\ region, corresponding to their emission and absorption layers. 
To model the CO $J$  = 8--7 and $J$  = 5--4 lines, we need a cloud in which the density in the foreground portion of the cloud is much lower than that of the ``hot'' PDR region producing the \OI\ emission.  
We have adopted a foreground density $n$(H) = 250 \cmv, which will largely be constituted by molecular hydrogen with $n$(H$_2$) $\simeq$ 125 \cmv.  
This value is adopted due to several constraints. 
First, too low a density will make the size of the low--density region (having $N$(H$_2$) $\simeq$10$^{22}$ \cmv) excessive compared to the scale of condensations seen in W3;
$n$(H$_2$) = 100 \cmv\ implies $L$ = 10$^{20}$ cm, or 32 pc, which is already very large.
Second, too high a density increases the CO formation rate and thus the CO abundance in the foreground region, reducing the abundance of atomic oxygen. 
Too low an atomic oxygen density in the foreground would make it impossible to obtain the values of optical depth and column density given in Table \ref{tab:fgcol}. 
A higher density also can produce some excitation of the $J$=4 level, which results in excessive absorption of the $J$=5-4 transition and line intensities much weaker than those observed.
Thus, the CO observations constrain the model of the cloud to have a very large, diffuse foreground, an appreciable fraction of the size of the entire W3 region.  

Consistent with the density constraints on the foreground, we have adopted a simple model in which we fix the total proton density ($n$(H) = 250 \cmv) and column density ($A_v$ = 10 mag.) of the foreground component. 
 We have adopted a somewhat enhanced cosmic ray ionization rate of 10$^{-16}$ s$^{-1}$ throughout, which has the effect of modestly reducing the fractional abundance of CO in the low--density region with a concomitant increase in the abundance of C$^+$.  

\begin{figure*}[htb!]
\includegraphics[angle=0, width=14 cm]{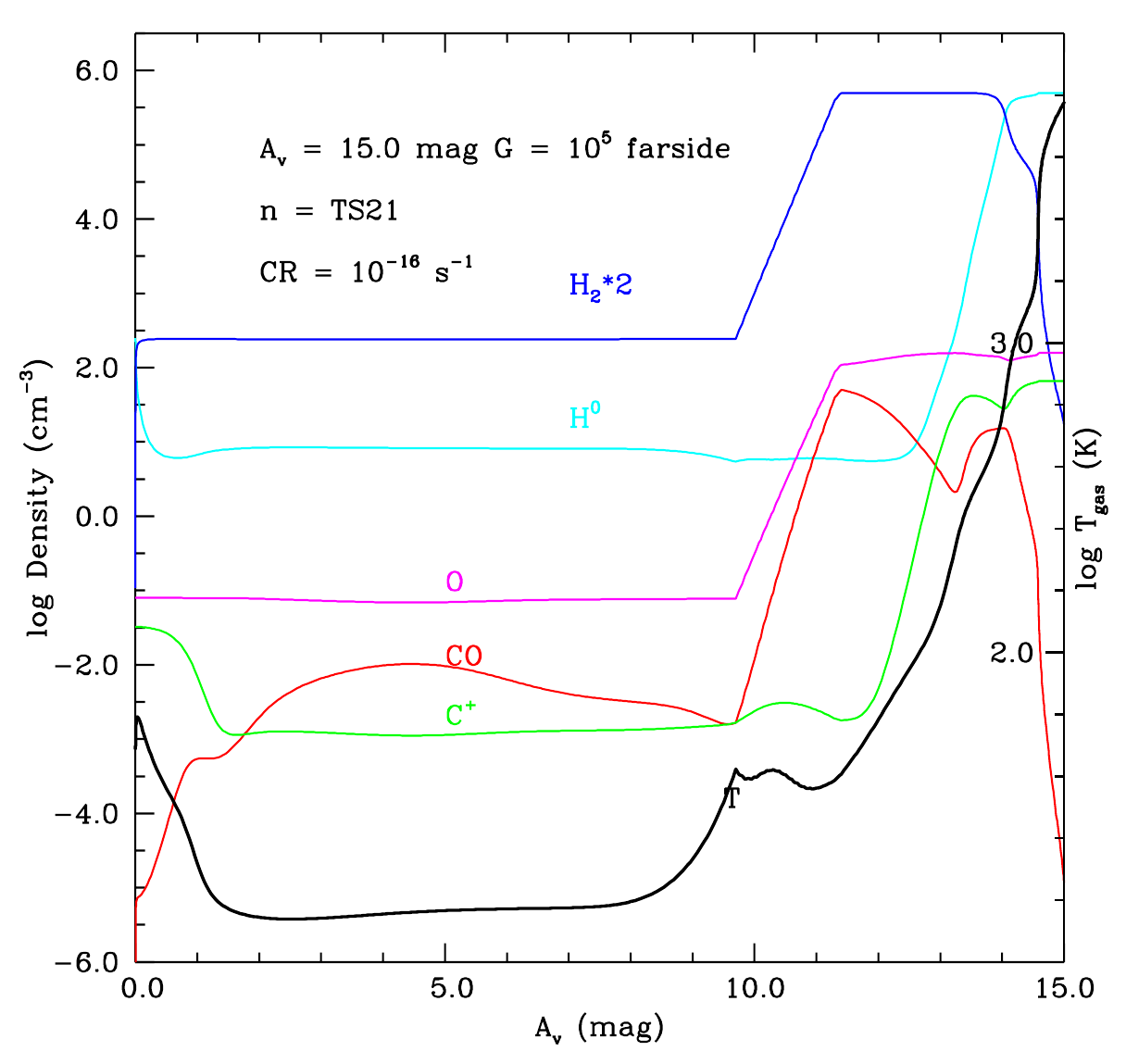}
\centering
\caption{Physical conditions and chemical abundances from the Meudon PDR code with parameters used for modeling W3.  The heating source, with intensity enhanced compared to the standard interstellar radiation field  (ISRF)by a factor of 10$^5$, is on the far side of the cloud (at $A_v$ = 15 mag.) and the density of the ``hot'' PDR region producing the fine structure lines and strong mid/high--$J$ CO line emission is 5$\times$10$^5$ \cmv\ for a region spanning $A_v$ = 3.6 mag.  At greater distances from the heating source, the density drops, and reaches a proton density $n$(H) = 250 \cmv\ at 5 mag. from the heating source, which is mostly H$_2$.  The thickness of the ``low density'' foreground cloud is $A_v$ = 10 mag.  The near side of the cloud is irradiated by the standard ISRF, which is responsible for the rise in temperature and C$^+$ abundance, and drop in CO abundance seen in the first 1--2 mag. of the cloud.  The transition from H$_2$ to H$^0$ takes place in the much thinner layer defined by 0.25 mag. total column density.  
}
\label{fig:PDR1} 
\end{figure*}

We have run the Meudon code for a range of input parameters.
Figure~\ref{fig:PDR1} shows the results for one such model density profile (denoted TS21), which is relatively successful in reproducing the observations.  
For this model,  the logarithm of the density in the transition from the low density foreground to the high density \OI\ emitting region rises linearly with extinction into the cloud over a thickness corresponding to $A_v$ = 1.5 mag. 
The density then remains at a fixed high value, $n$(H) = 5$\times$10$^5$ \cmv, until the heating source is reached.  
The total thickness, corresponding to $A_v$ = 15 mag., is determined by the extinction to the exciting stars and the \HII\ region W3A discussed above. 
The density of the high--density component cannot be less than 10$^5$ \cmv\ and still produce \OI\ intensities (more specifically, the high intensities derived from fitting the line wings and shoulders discussed above and presented in Table \ref{tab:fgcol}).

The hydrogen throughout the PDR is molecular with the exception of the two boundaries.
The larger of these regions, in which the hydrogen becomes atomic is 1 mag. thick, is located adjacent to the \HII\ region.  The second is a very thin region (barely visible in Figure~\ref{fig:PDR1}) where the standard ISRF produces a thin layer of H$^0$.   
The density and fractional abundance of atomic oxygen track the total proton density well, with only CO being another significant reservoir of oxygen, incorporating 15\% of that element.  
The CO abundance drops at both boundaries of the PDR region, where the carbon is converted to C$^+$.

The two CO lines observed constrain the high--density portion of the cloud in the framework of the two--component model we have developed.
In Figure~\ref{fig:co4pos} we show the spectra of the two CO transitions at four positions.  The three positions relatively close to the central position share the characteristics that (1) the CO $J$=5--4 line shows self-absorption, (2) the CO $J$=8--7 is single--peaked or shows only slight self--absorption, and (3) the maximum intensity of the $J$=8--7 line is modestly greater than the peak or peaks of the $J$= 5--4 line.  
 
At the fourth position, at the extreme Eastern end of the linear array of positions observed, neither line shows obvious self--absorption, but the two emission spectra show significantly different peak velocities.  
There is very considerable velocity structure within the W3 region, as discussed in \S\ref{sec:Velocity}, and the shift as well as the quite different line profiles may result from velocity gradients along the line of sight between the locations dominating the emission from the two different transitions.
The CO(8--7) line here is somewhat weaker than the lower--lying CO(5--4) transition, which could reflect the temperature gradient throughout the PDR.   
It is difficult to produce greater absorption in the $J$=8--7 line than that in the $J$=5--4 line, so the different profiles and peak velocities are likely the result of the velocity structure within the extended PDR, combined with the temperature gradient.  
Small differences in the velocities of the multiple components contributing to the profile observed at other positions  can be seen by careful examination of the spectra at the other positions, albeit somewhat confused by the strong absorption present in the $J$= 5--4 line.

\begin{figure*}[htb!]
\includegraphics[angle=0, width=14 cm]{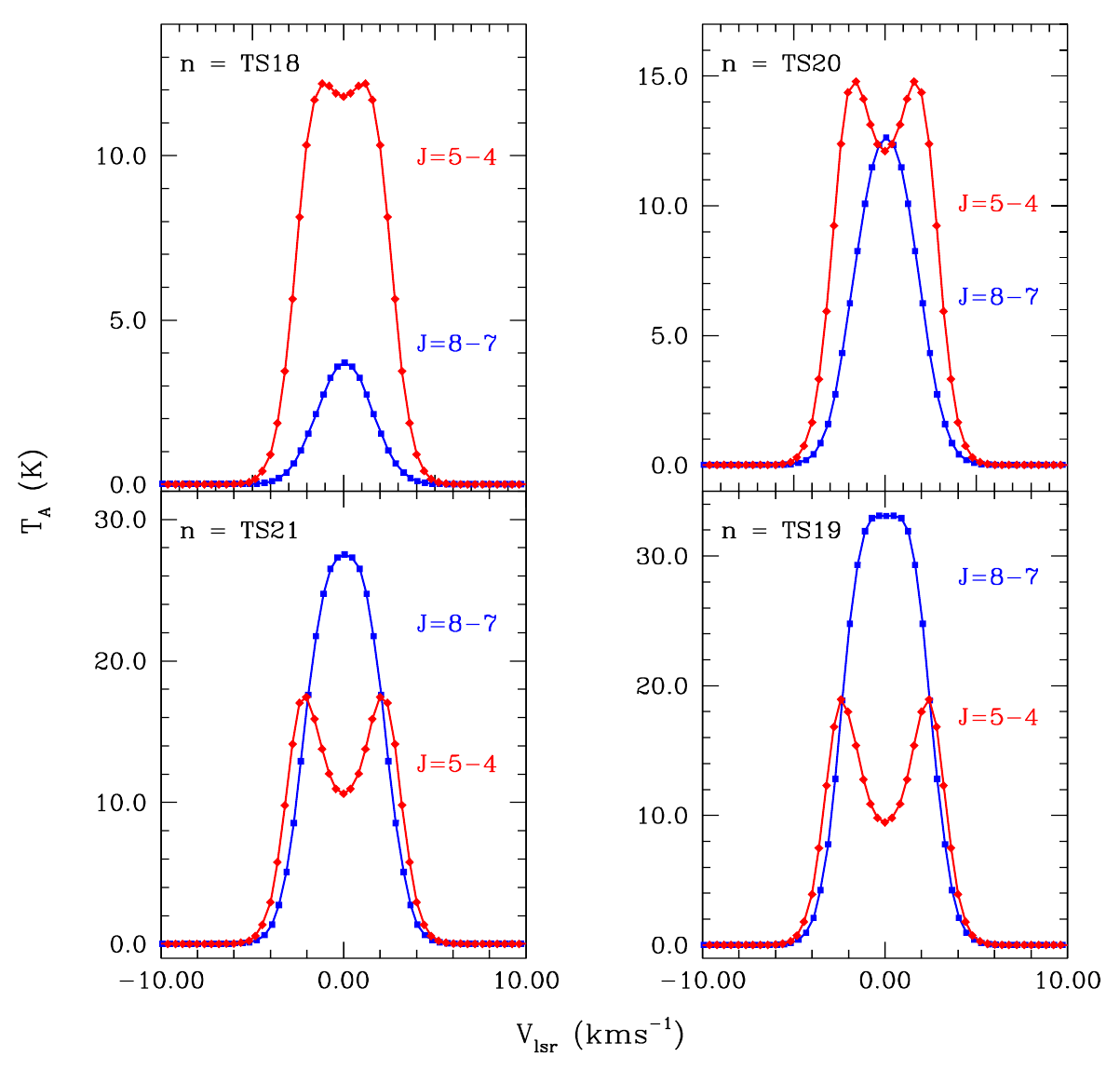}
\centering
\caption{\label{co_4models}  Spectra of CO $J$=8--7 (blue lines) and $J$=5--4 (red lines) for four different PDR/foreground cloud models.  All models have foreground proton density $n$(H) = 250 \cmv, and extent corresponding to $A_v$ = 10 mag.  The density rises to that of the PDR region in front of the heating source having $G$ = 10$^5$.  The cosmic ray ionization rate is 10$^{-16}$ s$^{-1}$ throughout.  The four models differ in the density of the high--density PDR portion of the cloud which have proton densities equal to 10$^5$, 2$\times$10$^5$, 5$\times$10$^5$, and 10$^6$ \cmv\ for TS18, TS20, TS21, and TS19, respectively.  
}
\end{figure*}

We have found that the density of the ``high density'' portion of the cloud has a major effect on the observed CO spectra. 
To quantify this, we have run models with different densities in this region, but keeping other parameters the same. 
In Figure \ref{co_4models}, we show the predicted line profiles for four ``tapered step'' models: TS18, TS19, TS20, and TS21.  
The proton densities in the high--density region are 10$^5$, 2$\times$10$^5$, 5$\times$10$^5$, and 10$^6$ \cmv, for models TS18, TS20, TS21, and TS19, respectively.  
We see a systematic increase in the intensity of the CO $J$=8--7 line with increasing density, with the line clearly becoming flat--topped due to saturation for the highest density.  
The intensity of the $J$=5--4 line increases modestly with increasing density, but the absorption gets stronger as well, being very small for model TS18 having density in the high--density region equal to 10$^5$ \cmv.
The intensity of the $J$ = 8--7 line increases rapidly with increasing density, but also becomes optically thick and flat--topped for high--density region density  equal to 10$^6$ \cmv.
To be consistent with the observations,  we require the peak $T_{mb}$(J=8--7) to be 20\% to 50\% stronger than the maxima of $J$ = 5--4, and the $J$ = 5--4 line to have an absorption dip of $\simeq$50\% relative to the residual maxima on each side of the dip.
None of the models quite satisfies this second requirement, but we are constrained here by the Meudon code requiring the same line width for absorbing and emitting regions. 
With the observed narrower absorption line width, the absorption in the model would be deeper and would agree reasonably well with the observations for a high--density region density of 5$\times$10$^5$ \cmv.
The TS21 model with this density is close to the best fit to the shared characteristics of the spectra observed at the three central positions of the strip.

The PDR models indicate that the 146 \um\ \OI\ emission is hardly affected by the change in density, being  only 2 K lower for the two lower--density models compared to the 34 --36 K peak values of two higher--density models.
The 63 \um\ emission is almost completely unaffected, with the residual ``horns'' resulting from the absorption dropping by 1 K from the 8 K found for the values of the three lower--density models.
The density--independence results from the populations of the fine structure levels being close to LTE in the high--density region for the higher--density models.  
Note that the formally--defined critical density for the 146 $\mu$m line is misleading in terms of the closeness to LTE populations due to the population inversion present over a significant range of densities \citep{Goldsmith19}.
Thus, while the \OI\ absorption does not put tight constraints on the properties of the foreground gas, the two CO lines do constrain them reasonably well, and we adopt the TS 21 model as the best representation of the hot, dense PDR and lower--density foreground material in W3.

The Meudon model does not allow for any changes in line width or central velocity as a function of position in the cloud.  
Thus we cannot expect to duplicate the details of the observed spectra.  But focusing on the shared characteristics of the three positions toward the center of the strip, we can see that model TS21 shows the best agreement in terms of line shapes and intensities, and in particular the absorption feature seen in the $J$=5--4 spectrum.

\subsection{Modeling \OI\ Emission}

The intensity of the background \OI\ emission is dependent on the temperature of the high--density PDR layer and thus the heating from the \HII\ region adjacent to it. 
In the following discussion we adopt $G_0$ as the standard Mathis interstellar radiation field intensity, and denote the source intensity in terms of $G\times G_0$, where $G$ is dimensionless.  
Considering a region without foreground absorption and $G$ = 10$^5$, the \OI\ peak antenna temperature, $T_A$(63 \um) = 80 K, while for $G$ = 10$^6$, $T_A$(63 \um) = 120 K.  
Thus to reproduce the strongest fitted background temperatures (Figure \ref{fig:OIspec}), we require a radiation field even higher than $G$ = 10$^5$. 
The requirement for a larger radiation field for these positions suggests that the stars responsible for producing the ionization are very close to the boundary of the PDR, while the other positions along the cut shown in Figure \ref{fig:OIspec} are at a somewhat greater distance from the stars providing heating and ionization.
 We have adopted $G$ = 10$^5$ as a representative value for analyzing the behavior of the foreground absorption.

\begin{figure*}[htb!]
\includegraphics[angle=0, width=14 cm]{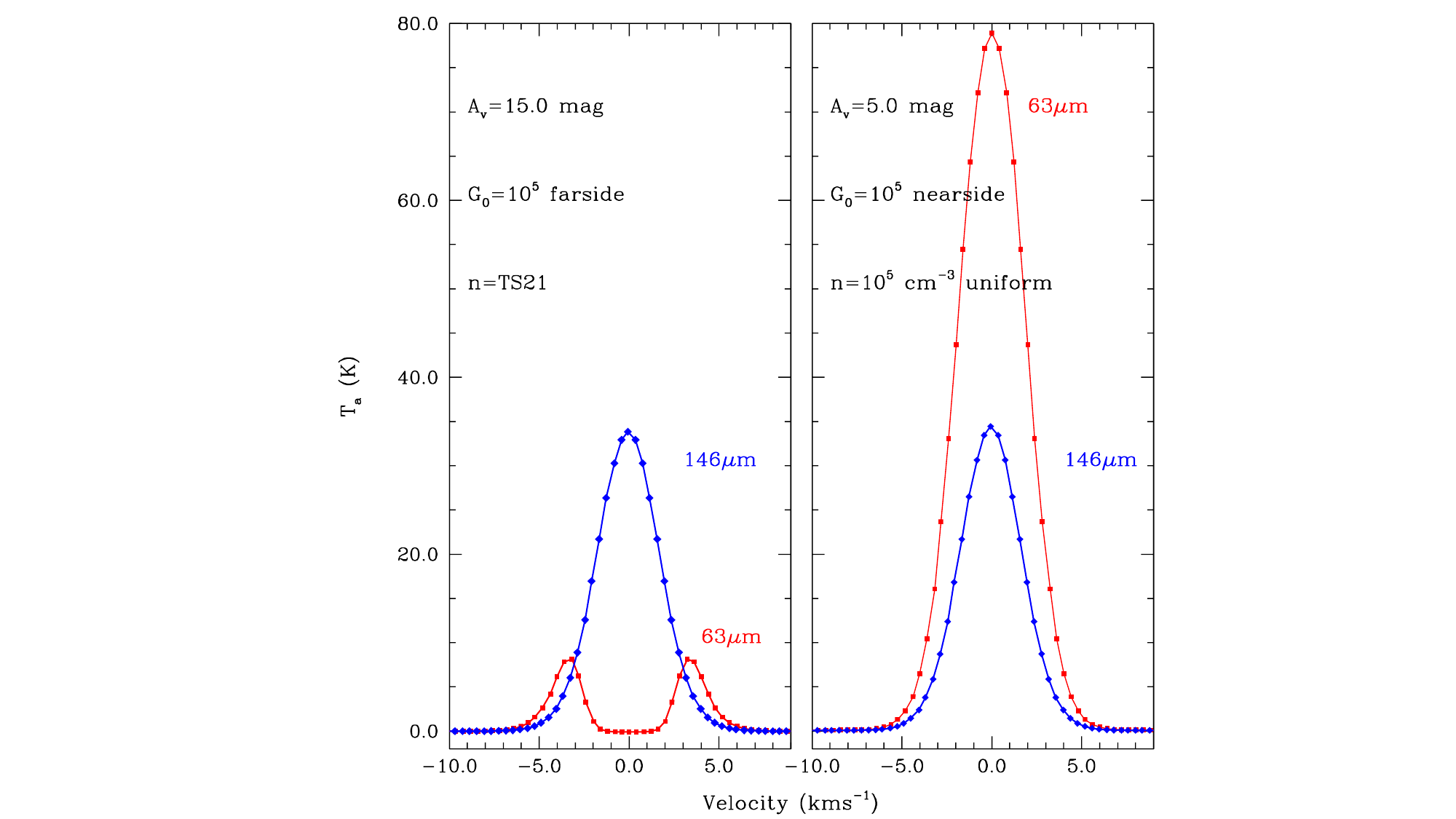}
\centering
\caption{Effect of foreground absorption on \OI\ emission spectra. The left hand panel shows spectra of 63 \um\ and 146  \um\ \OI\ lines predicted by the TS21 Meudon PDR code model, having total foreground and background densities $n$(H) = 250 \cmv\ and 5$\times$10$^5$ \cmv, respectively, with the heating source on the far side of the cloud.  The extent of the high--density region corresponds to approximately 5 mag., with $\simeq$ 4 mag. being at density 10$^5$ \cmv\ or higher. For comparison, the right hand panel shows the spectra if there were no intervening low--density gas in front of the hot PDR, assuming $n$(H) = 10$^5$ \cmv\ and a slab of column density similar to that of the high--density region. 
\label{fig:OIcomb}
}
\end{figure*}

The dramatic effect of the foreground material that is included in the TS21 density profile (Figure \ref{fig:PDR1}) is illustrated in Figure \ref{fig:OIcomb}. 
The left panel shows the result of including the complete density profile; the 63 \um\ line intensity is reduced to essentially zero over approximately the intrinsic FWHM line width. 
This result is very similar to that seen at positions (+14,-16), (+1,-2), and (-14,+13) in Figure \ref{fig:OIspec}, with the largest \OI\ optical depths in the foreground gas between 5.0 and 6.5. 

The observed line profiles shown in Fig. \ref{fig:OIspec} are all quite asymmetric as a result of  the velocity of the absorbing material being shifted relative to the background emitting oxygen.  
Velocity shifts on the order of 1 \kms\ are sufficient to produce the highly asymmetric emission observed at some positions, and this asymmetry is discussed further  in \S\ref{sec:Velocity}. 
Larger shifts result in essentially  single--peaked observed peak emission that is significantly shifted from the inferred centroid velocity of the background emission.  

The  agreement between observed and modeled line profiles would be even closer were it not for the slight velocity shift between background emission and foreground absorption.  We are unable to include this velocity shift  in the models as the Meudon code does not allow for a varying velocity profile.

For the \OI\ spectra the foreground line widths are slightly less than those of the background emission, which is not surprising given the greater turbulence in the hot PDR immediately adjacent to the heating source. 
In addition, the large 63 \um\ optical depth in the hot PDR contributes to the line broadening that is clearly seen in Figure~\ref{fig:OIspec}. 
The right panel in Figure~\ref{fig:OIcomb} shows the results of a model with 5 magnitudes of material at a density of 10$^5$ \cmv; very close to that of the portion of the TS21 model near the heating source.  
Here, the 63 \um\ line is not absorbed, but is broadened due to high optical depth. 
The peak antenna temperature is close to 80 K.  
The 146 \um\ line intensity is essentially independent of the relative location of hot and cooler material, and thus is unaffected by foreground absorption,  as this line is optically thin.
The observation of foreground \OI\ 63 \um\ absorption along a given line of sight is thus highly dependent on the geometry.
Observations of an external galaxy, for example,  in which multiple clouds are in the antenna beam will still suffer from the absorption reducing the measured flux.  

The Meudon PDR code also calculates the densities in each fine structure level at each position in the cloud. 
The H$_2$ density in the low--density foreground region is 121 \cmv, and its physical thickness is 8.3$\times$10$^{19}$ cm, or 26 pc.
Integrating the O$^0$ volume density through the foreground layer, we find the total atomic oxygen column density in the $^3$P$_2$ ground state to be 6.4$\times$10$^{18}$ \cmc.   
Taking a typical line width for the foreground gas of 5 \kms, we find a peak optical depth equal to 5.5. 
This value agrees well with those deduced from fitting the observed spectra, giving high confidence in the column densities in the foreground layer, although its volume density and physical extent are less certain.

\section{ \CII\ Spectral Analysis}
\label{sec:CII}

C$^+$ fine structure line emission was found by \citet{Howe91} to be extended in the W3 region, though located primarily to the south of W3A and W3 IRS5.
These observations had a velocity resolution between 67 \kms\ and 91 \kms\, so was unable to resolve line profiles or detect absorption if present.
Although we do not have \CII\ data for the positions at which we have observed \OI\ and CO, it is of interest to examine what the PDR model that we have developed predicts for \CII\ fine structure line emission and possible absorption.
Observations of other \HII\ region/GMC sources \citep{Guevara20, Jacob22, Kabanovic22} have revealed that there is significant self--absorption in \CII.

Three of the four sources studied by \citet{Guevara20} had foreground (absorbing) \CII\ column densities $\ge$ 10$^{18}$ \cmc, with the fourth having an order of magnitude smaller column density.   The peak absorption optical depths were between 1 and 2.
These authors had two major conclusions bearing on our modeling of \CII\ emission and absorption in W3.
The first is that no single slab of heated material could reproduce the observed emission, but that multiple ``stacked'' layers were required.
The second is that they could not explain the foreground column density and optical depth in any essentially uniform (unclumped) model.

\citet{Jacob22} reported \CII\ observations of two positions, W3(OH) and W3 IRS5, in W3, and one position in  NGC7538.  
Towards W3(OH) they find absorption column densities $N$(C$^+$) between $\simeq$ 5 and 23 $\times$10$^{17}$ \cmc, and for W3 IRS5, $N$(C$^+$) between $\simeq$ 6 and 8 $\times$10$^{17}$ \cmc. 
The velocity of strongest \CII\ absorption is -45 \kms\ in W3(OH) and -40 \kms\ in W3 IRS5.  
They also detect \OI\ 63 \um\ absorption at essentially the same velocities.
The well--known large--scale velocity gradients in W3 explain the velocity difference for W3 (OH) relative to W3A/W3 IRS5 which are separated by 0.3$\degree$ (10.5 pc).

\citet{Kabanovic22} studied the RCW120 source, modeling this source as an expanding ring, and observing a variety of molecules as well as \CII\ and \13CII.   
The spectra in a large fraction of the ring suggest self--absorption in the common isotope.
Using the two C$^+$  isotopologue lines, they derived \CII\ absorption optical depths $\simeq$ 1, with the ionized carbon being largely in a layer in which hydrogen is atomic.  

With the assumption that the absorbing gas in front of W3A extends to cover W3 IRS5, we have examined the \CII\ spectrum from \citet{Gerin15} obtained using {\it Herschel} in double beam switch (DBS) mode.  
We show this spectrum in Figure \ref{fig:CIIspectrum}, which has been scaled to display the single sideband (SSB) main beam temperature.  
A continuum level of 3.5 K, corresponding to a single sideband level of 7 K, has been removed.  
This intensity is somewhat greater than the 5.57 K reported by \citet{Jacob22} and is likely due to the smaller beam size of {\it Herschel} compared to that of SOFIA.
Because only one sideband of the continuum can be absorbed by the spectral line, the observed line profile drops to a level of $\simeq$3.5 K (before removal of the continuum) at the center of the absorption feature.

We derived a spectral profile of the \CII\ from the emission region by fitting a Gaussian to the line wings and shoulders (but ignoring the clear excess emission below -45 \kms\ and above -27 \kms) following the procedure described above for fitting the \OI\ lines.
Finally, the foreground absorption, represented by a purely absorbing Gaussian line profile, was fit to replicate the observed spectrum.
The result is reasonable fit to the emission profile, but an excellent good fit to the absorption feature, with maximum optical depth of 3.8 at a velocity of -40.5 \kms.
The velocity of the absorption seen in \CII\ is almost identical to that seen in \OI, while the line width is somewhat narrower  (as seen  in Table \ref{tab:fgcol}). 
{The narrower line width in \CII\ may be due in part to the fact that its optical depth is somewhat lower than that of \OI\ and is less broadened by saturation.

\begin{figure*}[htb!]
\includegraphics[angle=0, width=14 cm]{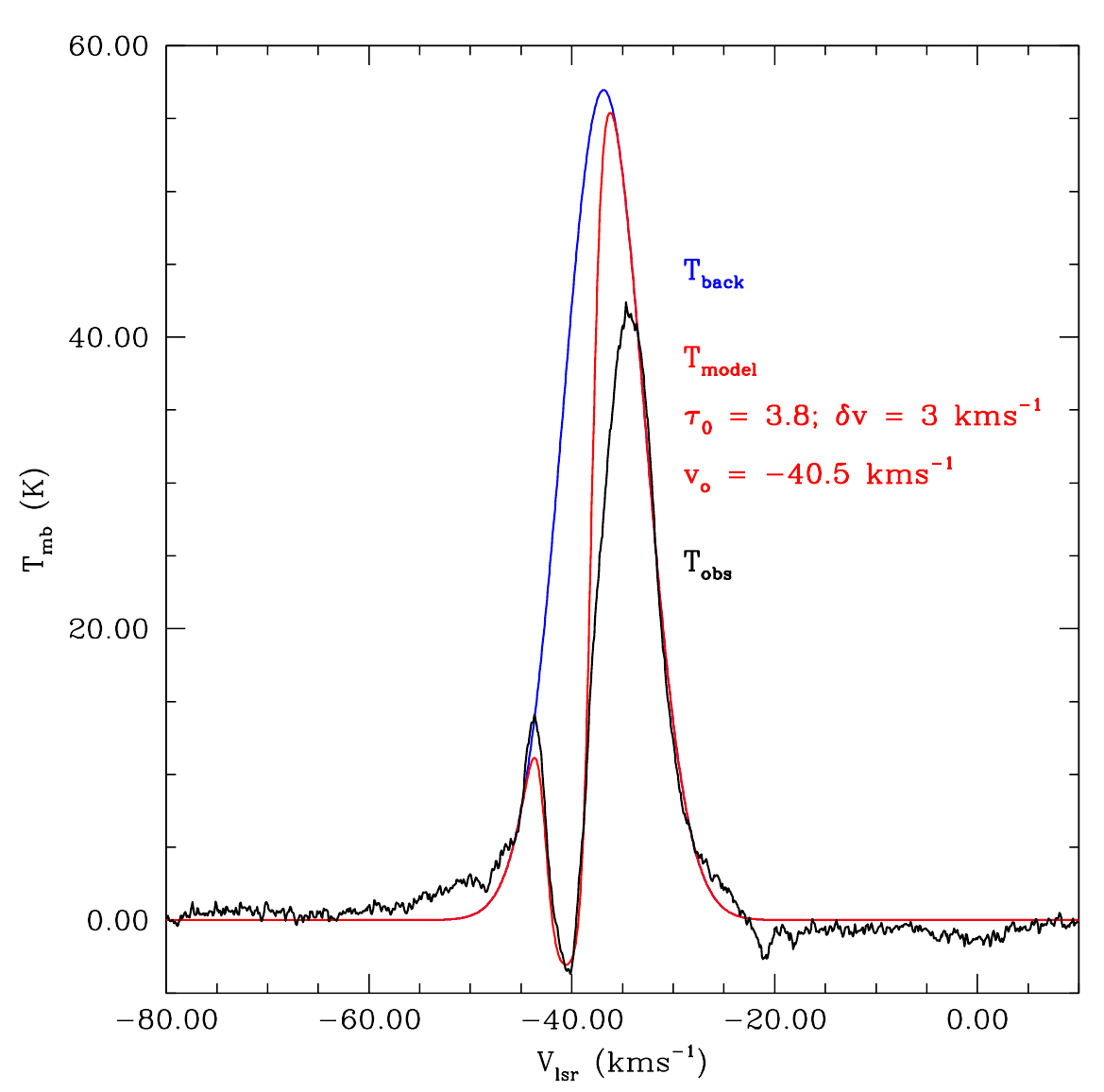}
\centering
\caption{Spectrum of \CII\ at W3 IRS5 from \citet{Gerin15}.  The data are shown in black, the fitted background emission profile in blue, and the complete model consisting of background plus fitted foreground absorption in red. The  peak absorption optical depth $\tau_0$ and centroid velocity $v_0$ of the absorption feature are also shown in red.
\label{fig:CIIspectrum}
}
\end{figure*} 	
Assuming the low--density foreground gas responsible for the \OI\ absorption is also producing the absorption feature seen in \CII, the density $n$(H) = 125 \cmv\ is much less than the critical density of 1.1$\times$10$^4$ \cmv\ derived using the deexcitation rate coefficients from \citet{Wiesenfeld14} (given that the hydrogen in this region is almost entirely molecular) and the spontaneous decay rate from \citet{Wiese07}.  
Thus, we can assume that all C$^+$ is in the ground state, and from Equation 35 of \citet{Goldsmith12}, we can write that
\begin{equation}
N({\rm C}^+)~({\rm cm^{-2}})= 1.4\times 10^{17} \int \tau(v) dv ~({\rm km s}^{-1}) .
\end{equation}
Approximating the integral by the fitted Gaussian profile yields $N$(C$^+$) = 1.7$\times$10$^{18}$ \cmc.
 This column density is consistent with that found by \citet{Jacob22} for W3 IRS5, and comparable to the larger values found by \citet{Guevara20}.  
 
 We find that the absorbing C$^+$ column density in model TS21 is only 1.8$\times$10$^{17}$ \cmc.  
 We endeavored to define a model with greater $N$(C$^+$) by decreasing the total proton density to 100 \cmv, while leaving other parameters unchanged.  
This has the effect of reducing the transformation of C$^+$ into CO, and with the result being $N$(C$^+$) = 6.9$\times$10$^{17}$ \cmc.
Although a significant increase, this model is still a factor $\ge$ 2 below what is required.
In addition the low density with the unchanged 15 mag. extinction makes the physical size of this region equal to 77 pc, which is considerably larger than  the entire W3 region.
Given that the cosmic ray ionization rate is already elevated at 10$^{-16}$ s$^{-1}$, it would seem very difficult to find a non--clumpy model with sufficient C$^+$ column density to match that indicated by the observations.
A clumpy model with penetrating UV radiation and C$^+$ on irradiated clump surfaces producing the observed \CII\ was proposed by \citet{Howe91} to explain the extent of the observed emission relative to the sources of ionization.
It is not obvious that this will resolve the present problem, as the interclump medium is generally taken to be quite warm, albeit of modest density,  Modeling with parameters constrained by observations of multiple molecular, atomic, and ionic species would be necessary to calculate whether this component or the ``envelopes'' of the denser clumps can provide sufficient absorption to explain the observations.
We thus confirm the conclusion of \citet{Guevara20} that more highly inhomogeneous models are likely required to reproduce the observed \CII\ absorption.

\section{Velocity Structure}
\label{sec:Velocity}

The W3 region has long been known to incorporate significant velocity differences due to the extended, highly inhomogeneous region and the star formation activity within it that has resulted in multiple regions of ionized and neutral gas.
Different tracers along a fixed line of sight also have different peak velocities; this is particularly true for a line of sight including ionized, highly--excited neutral gas, a PDR, and an extended foreground region.
We show the peak velocities of our data for \OI, CO, and \NII, in Figure \ref{fig:velocities} for 5 positions along the scan line shown in Figure \ref{fig:1.4GHz}.  

For the positions where we have been able to derive velocities for the absorption (in \OI\ and CO $J$=5--4) and emission, these are shown separately. 
In addition, we have velocities for the recombination lines H109$\alpha$ \citep{Rubin70}, C 90$\alpha$ \citep{Jaffe81}, and C166$\alpha$ \citep{R87}.  
The H 109$\alpha$ emission is expected to trace the fully ionized gas comprising the \HII\ region. 
The beam width used, $\simeq$6\am, is sufficiently large to encompass the entire region we have studied.
The \NII\  line traces similar material, so it is reasonable that the average velocity of the fine structure line is quite close to that of the hydrogen recombination line.  
Both are the most negative velocities found, $\simeq$ -41.5 \kms.
 
The C 90$\alpha$ recombination line (C RRL) observations (dashed black line) were made with a 90\as\ FWHM beam size while the C166$\alpha$ observations (dotted black line) had 14\as\ beam size. 
Due possibly in part to the significant velocity gradients within the W3 region, the two carbon recombination line velocities differ by almost 1 \kms.
The average carbon recombination line velocity, -39.4 \kms, is approximately 2 \kms\ red shifted relative to that of the fully ionized gas.

An early model for the carbon recombination lines, proposed by \cite{Zuckerman68},  is that they are produced at the interface between the fully ionized and the neutral gas.
Modeling including the temperature dependence of the C RRL showed that for moderately--large principal quantum number $n$, the primary temperature dependence is $T^{-1.5}$, favoring emission from cool gas \citep{Natta94}.
The \CII\ emission intensity, once the density is sufficient to thermalize the transition ($n$ $>$ $n_c$ = 4.5$\times$10$^3$ \cmv; \citet{Wiesenfeld14}), increases with temperature up to a few hundred K.
This increase is due simply to the increase in the fractional population of the upper level, which reaches an asymptotic value and does not increase further at higher temperatures.

C$^+$ is the dominant form of carbon for the portion of the PDR at $A_v$ $\leq$ 2 mag from the surface, {\it i.e.} in the range $A_v$ 13--15 mag in the PDR model (Fig. \ref{fig:PDR1}). 
This location, together with the temperature dependence of the emission, suggests that it arises very close to the interface with the \HII\ region, and is sensitive to the column density of C$^+$ at temperatures $\geq$ 150 K.  

Carbon recombination line emission, on the other hand, will be greatest from gas at much lower temperatures, as long as there is sufficient carbon there.  
In our PDR model TS21, the atomic carbon has a very clear peak at extinctions between 1 and 4 mag. from the \HII\ region, while the temperature falls to 100 K only for $A_v$ = 2 mag, and reaches 40 K at $A_v$ = 4 mag.
The Meudon PDR code does not calculate recombination line emission explicitly, but using the same code, \citet{Salas19} showed that C RRL emission for a wide range of $n$ peaks in exactly the  $A_v$ = 2--4 mag interval.  
These authors also indicate that this conclusion is not significantly affected by an increase in density and radiation field enhancement factor by a factor of 10 from their values of 10$^4$ \cmv\ and 10$^4$, respectively, and thus their conclusions would apply to the model with values we have adopted.

We thus expect the C RRL emission to peak in a region which is significantly removed from the \HII\ region interface.  
This offset is similar to that of the region responsible for CO $J$ = 5--4 emission, which has an upper level $E_u/k$ = 83 K above ground state.  
The CO $J$ = 5--4 emission has a velocity of -38.5 \kms\ at positions 1-3 (Southeast and central positions), but exhibits a very substantial gradient moving to the Northwest (positions 4 and 5).  

CO $J$ = 5--4 absorption arises from any cooler gas, but which has to be warm enough to populate the $J$ = 4 level, 55 K above the ground state.  
As can be seen in Figure \ref{fig:PDR1}, the CO abundance is dropping rapidly for extinction $>$ 3 mag. from the boundary of the PDR, and the gas temperature falls to 63 K at $A_v$ = 5 mag. from the boundary, where the CO density has fallen to a relatively low value due to the drop in the H$_2$ density.  
The region of CO $J$ = 5--4 absorption is thus only slightly displaced from that of the emission in this line. 
The CO $J$ = 5--4 absorption velocity is quite similar at the three positions were it could be clearly detected, with $v$ = -38.7 \kms.
This relationship  is quite similar to the \OI\ absorption, which has a varying velocity, with average value $v$ = -40 \kms\ for positions 1-3, and following the trend to more negative velocities as one moves to the Northwest.
The velocities of these molecular features are all quite similar to those of the C RRL, consistent with the picture discussed above.  

The \OI\ emission (like that of \CII) is dominated by the hot gas very near the \HII\ region.  
This feature is clearly red shifted relative to both the fully ionized and the general PDR material, with an average velocity $\simeq$-37.5 \kms.
The relative red shift of the \OI\ emission is striking, as is the fact that it is systematically offset in velocity from the foreground \OI\ absorption, as well as the velocities of other tracers of the extended PDR.

The blue shifted velocity of the fully ionized gas relative to the general PDR material could reflect a velocity gradient within the cloud.
It is possible that the PDR region has been accelerated toward us by the \HII\ region, although this is difficult to determine.
Since the relative red shift of the \OI\ emission is presumably a reflection of the motion of the surface layer of the PDR, which is on the far side of the entire PDR as seen from the Earth, it implies that the very hot gas has been accelerated away from the bulk of the PDR   (in this case from the Earth). 
This type of flow away from the ionized--neutral boundary is reminiscent of that proposed for the Orion HII region by  \citet{Zuckerman73}.

\begin{figure*}[ht]
\includegraphics[angle=0, width=14cm]{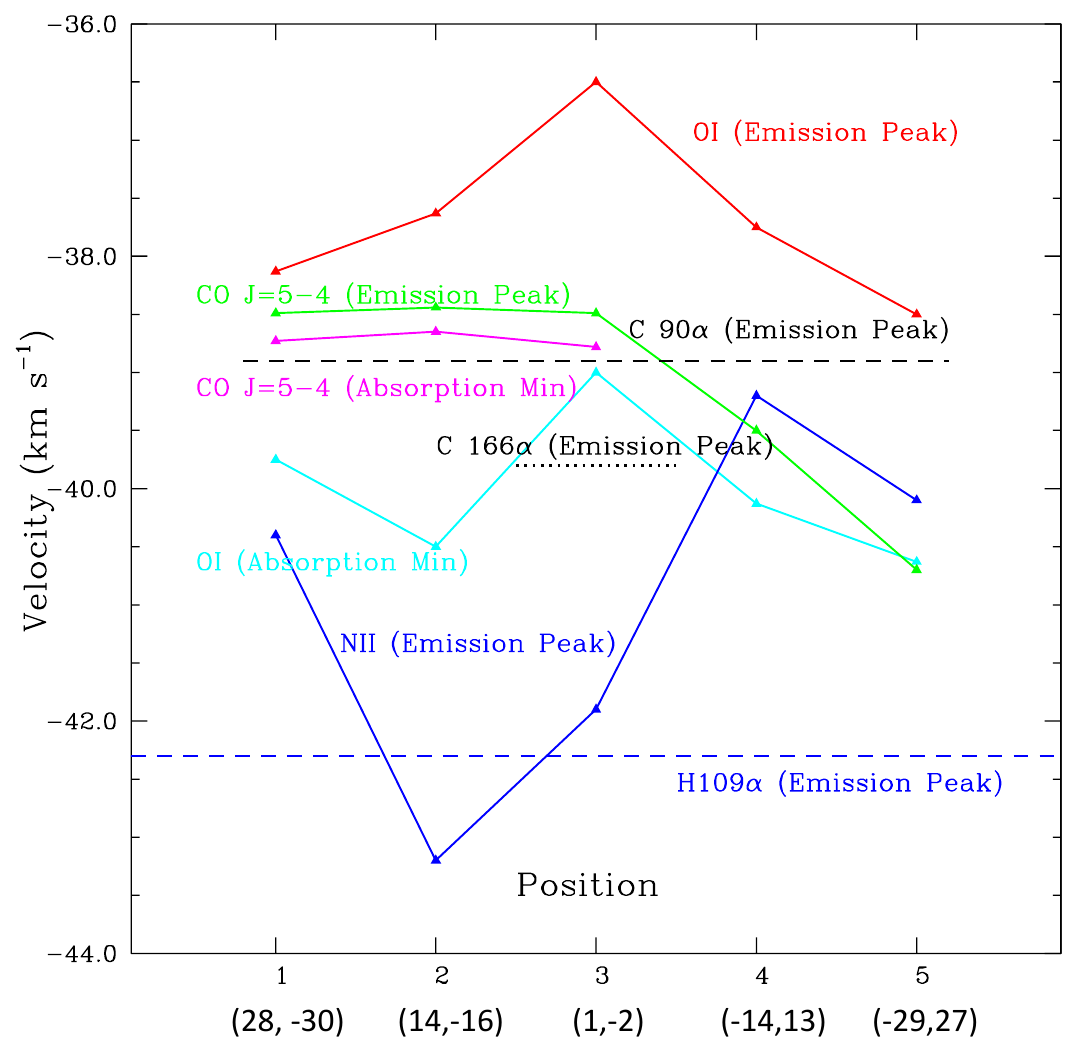}
\centering
\caption{Velocities of tracers of different components of W3 \HII\ region/PDR/ foreground cloud.  Emission peaks from fitted Gaussians and absorption minima for \OI\ 63 \um\ and CO $J$ = 5--4 are shown separately.  The points shown are limited to the positions at which reliable detections were made.  The positions are numbered from Southeast to Northwest and the offsets in seconds of arc from the central position are also given.  The H109$\alpha$ radio recombination line is a large--beam result \citep{Rubin70} that encompasses  the entire region. The C90$\alpha$ observations (dashed black line) were made with a 90\as\ FWHM beam size. The C166$\alpha$ observations (dotted black line) had 14\as\ beam size.  The peak velocities of the two carbon recombination lines differ by almost 1 \kms.  The velocities of the tracers of the fully ionized gas (H109$\alpha$ and \NII) differ systematically from those of the material in the PDR (\OI\ absorption, CO emission and absorption, and carbon recombination lines).  The \OI\ emission is significantly blue shifted relative to the other PDR lines.
\label{fig:velocities}
}
\end{figure*} 	

\section{Discussion}
\label{sec:Discussion}

The observations and modeling presented above confirm the presence of significant foreground absorption over a $\simeq$ 2 pc portion of the W3 cloud, resulting in a reduction in the observed \OI\ 63 \um\ line intensity by a factor of 3 to 5.  
The more comprehensive analysis of the emission from different positions along the linear cut made through W3A reveals that the absorbing layer is quite extended in the plane of the sky, so must be regarded as a foreground feature that regularly accompanies gas heated by the \HII\ region that is responsible for \OI\ fine structure line emission.  
This confirms the importance of geometry in determining the presence of absorption and of the reduction in observed total \OI\ 63 \um\ line flux.

\OI\ self--absorption was predicted in the modeling of \citet{Vasta10}, attempting to explain the intensities of \OI\ 63 \um\ relative to those of other fine structure lines. 
The present work extends and quantifies the analysis of foreground absorption which is related to, but not the same as, optically thick \OI\  \citep[e.g.,][]{Stacey83,Stacey93}, observations that could not directly detect absorption features.

The absorption seen in mid--$J$ CO transitions helps define the transition from the high--density, hot component of the PDR near the \HII\ region to the much more physically extended cooler component.  
These lines are valuable complements to the 63 \um\ \OI\ fine structure line, which, having a much higher equivalent temperature ($T^* = hf/k)$, and critical density, is absorbing over a wide range of densities and temperatures and unable to provide very tight constraints on these parameters in the foreground absorbing layers.

The present observations did not include observations of the \CII\ 158 \um\ fine structure line, but other observations have reported significant optical depth and foreground absorption in this transition at other nearby positions in W3  \citep{Guevara20,Jacob22, Kabanovic22}.  
Assuming that  the column density of C$^+$ producing the absorption observed is similar to that seen at the positions modeled here, we find that our PDR model cannot reproduce  the required absorbing column density, confirming the results of \citet{Guevara20} for other sources.  
This disagreement could be a suggestion that small--scale density non-uniformities  (``clumps'') may be present in the extended cloud material associated with PDRs and \HII\ regions.

\section{Summary}
\label{sec:Summary}

We have presented a detailed model for the structure of the gas producing strong FIR line emission and absorption in the W3A \HII\  region.
The primary purpose is to understand better the strong absorption seen in the 63 \um\ atomic oxygen fine structure line.  
We have used a combination of fits to the wings and shoulders of the emission profile and of the absorption feature at 6 positions to determine the column densities and optical depths of the low--excitation atomic oxygen responsible for the absorption.  
Observations of the $J$ = 8--7 and $J$ = 5--4 transitions of CO show prominent self--absorption in the lower transition in the central portion of a strip map through W3A.  

We have developed a two--component cloud model to explain our observations in which an extended, low--density region lies in front of the  high--density, hot PDR adjacent to the \HII\ region, which is powered by massive young stars.  
We have combined this model with the Meudon PDR code to determine the physical conditions and chemical abundances, and to produce line profiles for the oxygen and carbon monoxide lines observed.
The total extinction derived to the exciting stars fixes the total column density of the observed system, while the production of the strong emission derived for the background PDR requires strong heating (G = 10$^5$) and high densities.
The density is significantly constrained by the CO observations with best--fit model having $n$(H) = 5$\times$10$^5$ \cmv.  

The absorption by low--excitation atomic oxygen determines the column density of the foreground gas, and we find a very good agreement between model and observations for total foreground hydrogen column densities $N$(H) =2$\times$10$^{22}$ \cmc.  
There are modest constraints on the density of the foreground gas, with a proton density $n$(H) = 250 \cmv\ favored, although this results in  the foreground material being spatially extended, with size $\simeq$26 pc.
The different spectral lines observed at a given position have different centroid velocities resulting from the complex kinematics of the gas being dramatically affected by the formation of massive stars.  

The observed velocities for carbon recombination lines agree with those of the CO $J$ = 8--7 emission, consistent with the modeling by \citet{Salas19} showing that the recombination line emission peaks at visual  extinctions of 3--4 mag. for \HII\ region/PDR parameters such as those describing W3A.

Our modeling confirms the presence of significant foreground absorption over a portion of the W3 cloud, $\simeq$ 2 pc in size, resulting in a reduction in the observed 63 \um\ line intensity by a factor of 3 to 5.  
This confirms the importance of geometry in determining the presence of absorption and of the reduction in observed total \OI\ 63 \um\ line flux.

\OI\ self--absorption was previously predicted in  attempting to explain the intensities of \OI\ 63 \um\ relative to those of other fine structure lines.
The foreground absorption is related to, but not the same as, optically thick \OI\  inferred from observations that could not directly detect absorption features \citep[e.g.,][]{Stacey83, Stacey93}.
\OI\ absorption  has not previously been analyzed over  such an extended portion of a massive star--forming region.  
The analysis presented here, using gaussian fits to the line wings and shoulders of absorbed spectra and a more realistic cloud geometry, shows that spectrally resolved \OI\  can be used as a tracer of star  formation even in the presence of absorption. 
\acknowledgements
We thank Dr. Franck LePetit for assistance and advice regarding the Meudon PDR code and Dr. Maryvonne Gerin for providing \CII\ data in W3 IRS5.  
We are grateful to the reviewer for many comments which clarified our presentation and impelled a more detailed discussion of uncertainties in Gaussian fitting and resulting optical depth determinations.
This research is based in part on observations made with the NASA/DLR Stratospheric Observatory for Infrared Astronomy (SOFIA).  SOFIA was jointly operated by the Universities Space Research Association, Inc. (USRA), under NASA contract NNA17BF53C, and the Deutsches SOFIA Institut (DSI) under DLR contract 50 OK 0901 to the University of Stuttgart. 
The contribution to this work by J. Stutzki, C. Guevara, R. Aladro, and M. Justen was supported through the Collaborative Research Centre 956, sub-projects A4 and D2 (project ID 184018867), funded by the Deutsche Forschungsgemeinschaft (DFG).
This work was performed at the Jet Propulsion Laboratory, California Institute of Technology, under contract with the National Aeronautics and Space Administration. {\copyright}2023 California Institute of Technology. USA Government sponsorship acknowledged.\\
%
%
\appendix
\label{sec:GaussFit}

\section{Gaussian Fitting Procedure and Uncertainties}
%
%

The  \OI\ emission detected in the W3 survey \citep{Goldsmith21} arises from hot and dense gas associated with the portion of the PDR closest to the \HII\ region. 
As can be seen in Figure \ref{fig:PDR1}, the temperature drops from several thousand K to several hundred K within 1 mag. from the boundary of the PDR, and within another magnitude drops below 100 K -- too cold to emit in the \OI\ 63 \um\  line.
 The cooler and less dense gas in the portion of the PDR yet further from the \HII\ region absorbs some of this emission, with the effect being strongest in the lines arising from transitions to the ground state. 
Absorption is strongest in \OI\ and to a lesser extent the CO lines, with CO(5-4) showing significant absorption and CO(8-7) having the least, if any, absorption.  
In order to derive the properties of the absorption region we need to compare the observed spectra with those that would be observed from the emission region.  
 
We have assumed that the emission region produces an \OI\  emission line with a Gaussian profile.  
The observed \OI\ spectra can be used to estimate the properties of the emission spectra because the absorption is weak outside the line's center.  
For the \OI\ spectra in W3, the absorbing gas is shifted in velocity from that of the emitting gas and has a smaller velocity dispersion. 
Under these conditions the unattenuated, or slightly attenuated, outer portions of the \OI\ lines can be used to reconstruct the intensity and line shape by fitting a Gaussian to the wings (see discussion by \citet{Al-Nahhal19} on Gaussian fitting algorithms). 
Due to the velocity shift, one wing is characteristically much stronger than the other, and the fitting process has to deal with this asymmetry.
The fit is made by blocking the velocity range that includes most of the absorption, and varying this range  stepwise until a best fit is derived, as demonstrated below using model spectra.  
Given the differing characteristics of emission and absorbing regions, we are guaranteed that a portion of the wings are essentially unattenuated, enabling the Gaussian fit approach.   

To demonstrate this approach  we have fit Gaussians\footnote{Gaussian fits were performed using KaleidaGraph version 4.54, Synergy Software, Reading, PA, USA.} to \OI\ lines generated from model calculations of \OI\ emission and absorption in W3 using the Meudon code. The Gaussians are of the form \cite[see][]{Goldsmith21},}

\begin{equation}
T(V)=T_0+T_p e^{-(V-V_p)^2/(\delta V)^2},
\end{equation}
where in Tables \ref{tab:W3_all_gaussfits} and \ref{tab:fgcol}, FWHM=1.665$\delta V$.\\

\section{Enhanced Model for More Realistic Spectral Line Modeling}

To make this procedure more realistic, we have developed a program that overcomes the limitation of the Meudon PDR code in that it allows for different portions of the cloud to have different velocities and line widths.  
This update is implemented using the level populations produced by the Meudon PDR code, but solving the radiative transfer through the cloud on a slab-by-slab basis with the specified line velocity and width at each position in the cloud.  
This does introduce some formal inconsistency, because the effect of a changing line center velocity reduces the peak optical depth and changes in the line width affect the absorption coefficient and optical depth as 1/$\delta$v, relative to the uniform parameters throughout the cloud employed by the original Meudon code.
However for the \OI\ line most clearly, but also for the CO lines and the \CII\ line, the densities in the foreground region are so far below the critical densities that the effect of trapping with maximum optical depth equal to few to 10 is modest.  
Any changes in that trapping due to adopting different line center velocities and line widths will thus be second order and can be neglected.

We first present results from model clouds having total extinction of 15 mag., illuminated by radiation field $G$=10$^5$.  
The total hydrogen density ($n$(H$^0$) + 2$n$(H$_2$)) distribution is the same as the TS21 model shown in Figure \ref{fig:PDR1} with the high--density portion adjacent to the boundary with the \HII\ region heating the PDR.
The additional parameters are that the line center velocity and FWHM line width are 0 \kms\ and 4 \kms, respectively, within 3.8 mag. of the heated--side boundary, -1.15 \kms\ and 3.25 \kms\ in the region between 3.8 mag. and 5.3 mag., and -2.3 \kms\ and 2.5 \kms\ for extinctions greater than 5.3 mag. from the boundary. 
These values are based on the spectra observed in the central 3 positions shown in Fig. \ref{fig:OIspec}.  
 
\begin{figure*}[htb!]
\includegraphics[angle=0, width=8 cm]{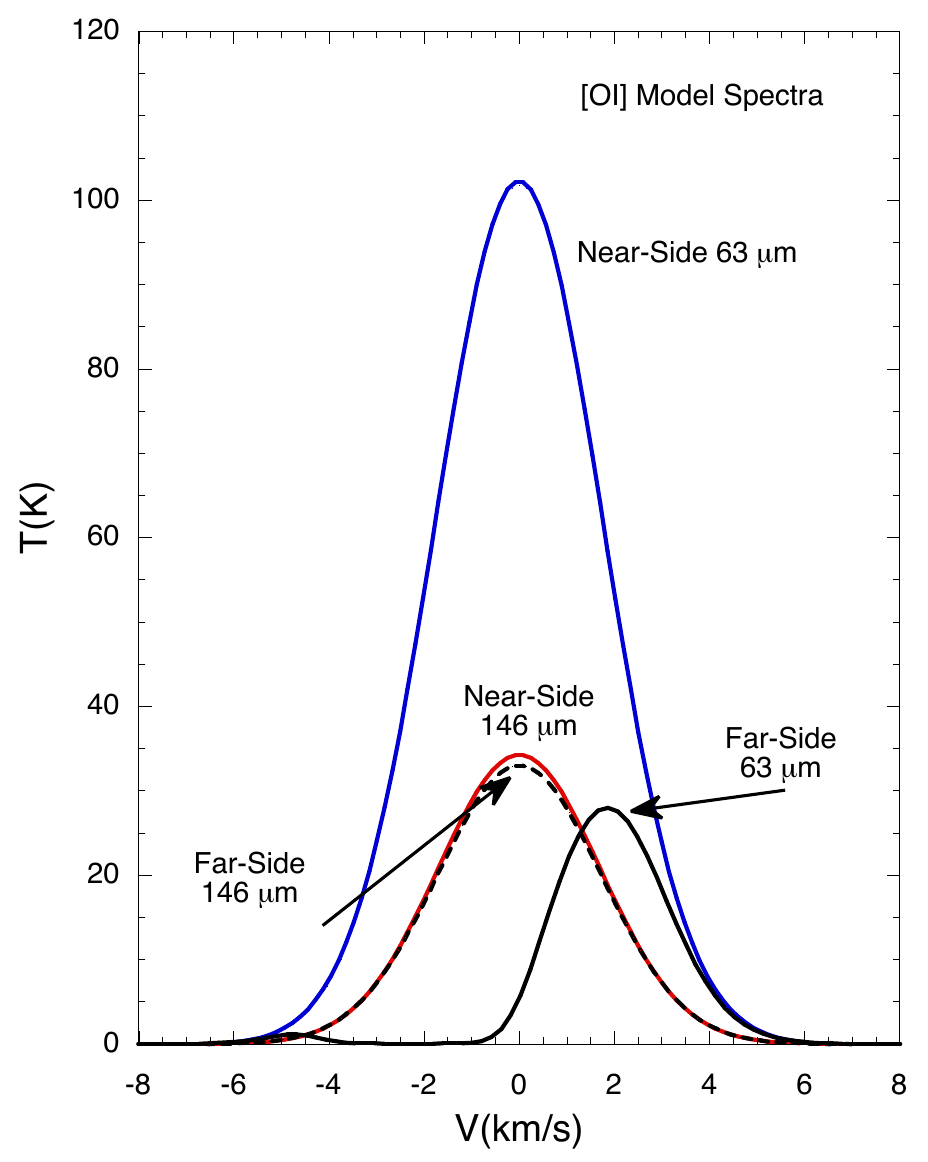}
\centering
\caption{\OI\ 63  \um\ and 146  \um\ spectra from Meudon PDR code with external radiative transfer calculation allowing line center velocity and line width to vary with position (see text).  The total extinction is $A_v$ = 15 mag. and heated boundary radiation field $G$ = 10$^5$.  The density profile is that shown in Figure \ref{fig:PDR1} with high density region always adjacent to the heated boundary.  Results from two configurations,  near--side (heated boundary closer to observer than low--density background cloud)  and far--side (heated boundary behind low--density foreground cloud), are shown.  The 146 \um\ line (red and dashed black curves) is virtually the same in both configurations, as it is unattenuated by absorbing oxygen atoms in the far--side configuration. Quite differently, the 63 \um\  line (blue and solid black curves) is heavily absorbed in the far--side configuration.   
\label{fig:FigA1}
}
\end{figure*}

Figure~\ref{fig:FigA1} shows the resulting spectra for the 63 \um\ and 146 \um\  lines that would be observed for two configurations:  (1) if the hottest portion of the PDR were facing the observer (``near--side source'') and (2) as observed with the HII region heating the PDR on the distant boundary of the source (``far--side source'').  
As expected, the peak intensity and profile of the 146 \um\ line are independent of the configuration, as this transition is essentially unattenuated due to the small population in the first excited state of \OI\ due to the low temperature and density of the region away from the heated boundary.  
In contrast, the 63 \um\ line is highly dependent on the configuration:  for the far--side configuration it is heavily absorbed as most of the \OI\ in the cooler foreground layer is in the ground state.  
In the near--side configuration, the low--excitation region is further away from the observer than the hot, dense gas, and so has little effect on the observed line.

\section{Gaussian Fitting of Near--Side Model to Derive Far--Side Spectrum}

A critical question is whether fitting a Gaussian to the wings of a line as observed in W3, which is a far--side configuration, gives the correct peak value of the emission that would be observed in the absence of foreground absorption.  
The best way to assess this issue is to compare the fitted Gaussian with the near--side spectrum that by definition includes the same heated region but with no absorption since it directly faces the observer.
To recover the ``unabsorbed'' 63 \um\ far--side line we fit the wings of the 63 \um\ far--side line with a Gaussian.
As we cannot know in advance the velocity range that is essentially unabsorbed we have to vary the velocity range stepwise to get a  best fit, as defined by the maximum $T_p$(K) generated by this process, while keeping the fitting error small (for the model Gaussians below the error is less than 2\%). 

\begin{figure*}[htb!]
\includegraphics[angle=0, width=16 cm]{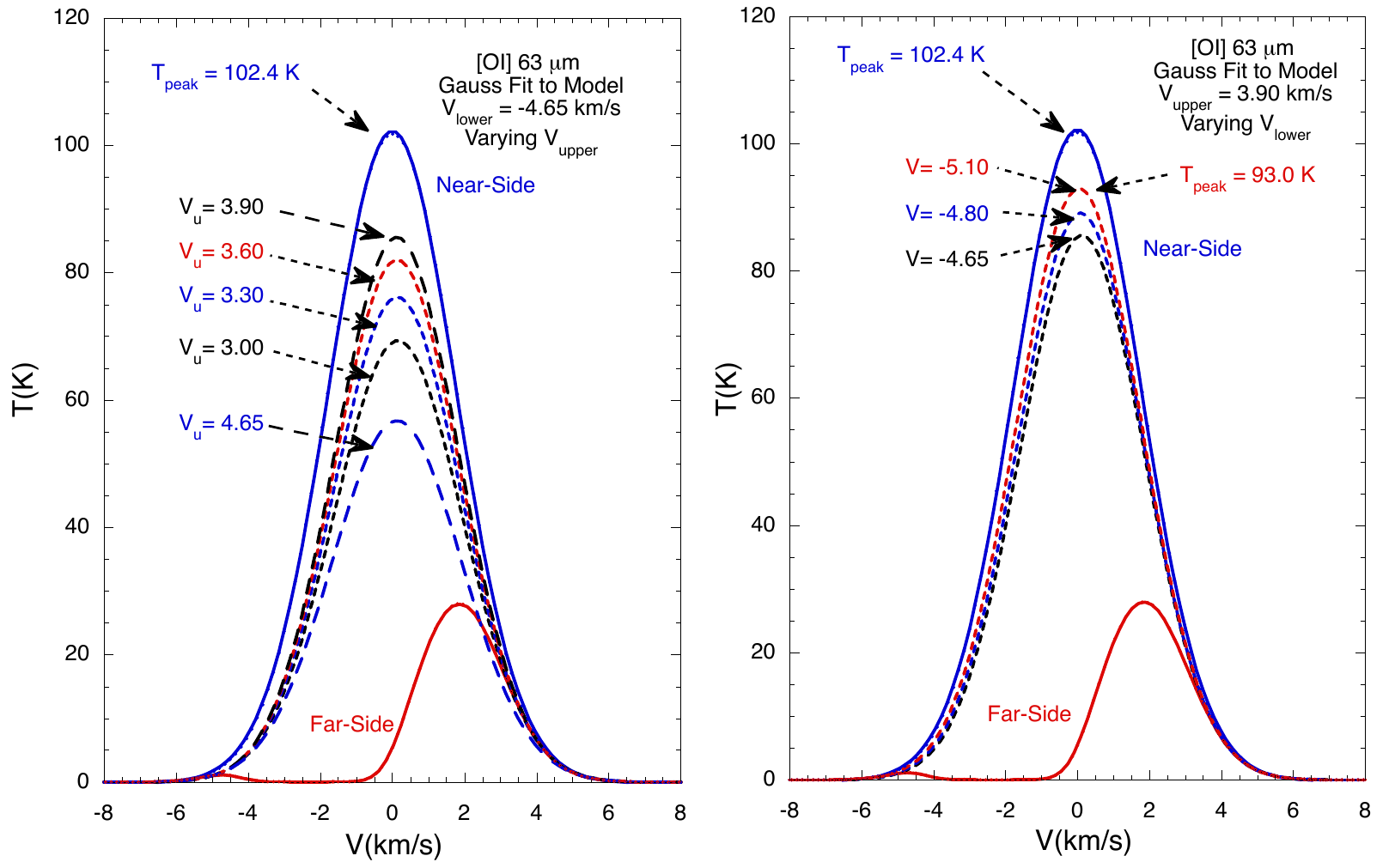}
\centering
\caption{The solid lines are the \OI\ 63 \um\ spectra shown in Figure~\ref{fig:FigA1} for the near--side (blue) and far--side (red) cloud configurations of cloud model described by Figure \ref{fig:PDR1} with spectra shown in Figure \ref{fig:FigA1}.  {\bf (left)} The dashed curves are Gaussian fits to the wings of the near-side 63 \um\ spectrum stepping through different blocked off regions in the high velocity wing of the line.  The peak intensity of the Gaussian fits at first increases with increasing blocking of the high velocity wing ($V_{upper}$ increasing from 3.00 to 3.90 \kms), but eventually decreases, as too much of the inner wing is removed from the fit, as shown for $V_{upper}$ = 4.65 \kms. {\bf (right)} Varying the blocking of the lower velocity ($V_{lower}$) wing, with $V_{upper}$ set to the value that maximized the peak intensity in the left panel. The dashed lines show how the Gaussian fits to the spectrum change as the lower velocity limit of the blocked part of the band is varied with the upper limit held fixed at the value that maximized the peak intensity in the left panel. 
\label{fig:FigA2}
}
\end{figure*}

In Figure~\ref{fig:FigA2} we show an example of the behavior of the fitted Gaussian line as the velocity range is varied stepwise.  
In the left panel the velocity of the lower boundary ($V_{l}$ is fixed at -4.65 \kms, and the upper velocity boundary of the blocked emission, $V_{u}$ is varied stepwise.  
As $V_u$ increases from 3.00 to 3.90 \kms\ the Gaussian peak increases to a maximum value of $\sim$86 K, within 84\% of the near-side peak emission of $\sim$102 K. 
Any further increases in the blocked velocity range reduces the projected near--side peak intensity, as shown by the Gaussian for $V_{u}$ = 4.65 \kms.

This behavior is readily explained by the nature of the absorbed emission as a function of velocity. 
If we start at the smallest blocked velocity, there are too many points that represent the absorbed region of the spectrum and the peak intensity is underestimated.  
As $V_{u}$ increases, the number of absorbed data points is reduced and more of the unabsorbed (or slightly absorbed) data points are included to weight the solution for the fitted Gaussian.  
At some point, an optimal number of data points is used for the fit and the amplitude of the solution peaks. 
Beyond that point, the data used for the fit is restricted to the more extreme the line wings, with the result that there is insufficient information about the curvature of the spectrum and the peak decreases and$/$or the solution becomes unstable.  

Once the best value of $V_{u}$ is known, the lower velocity range is varied in the same manner until the peak intensity is maximized.  
This process is illustrated in right panel of Figure~\ref{fig:FigA2} where it can be seen that the final (best) fit Gaussian has a peak of 93K, or $\sim$ 93\% of the near-side 63 \um\ emission.  
Throughout this iterative process  the central velocity and velocity dispersion hardly vary as they are constrained mainly by the symmetry and velocity range of the wings.


\section{Uncertainties}

\begin{figure*}[htb!]
\includegraphics[angle=0, width=8 cm]{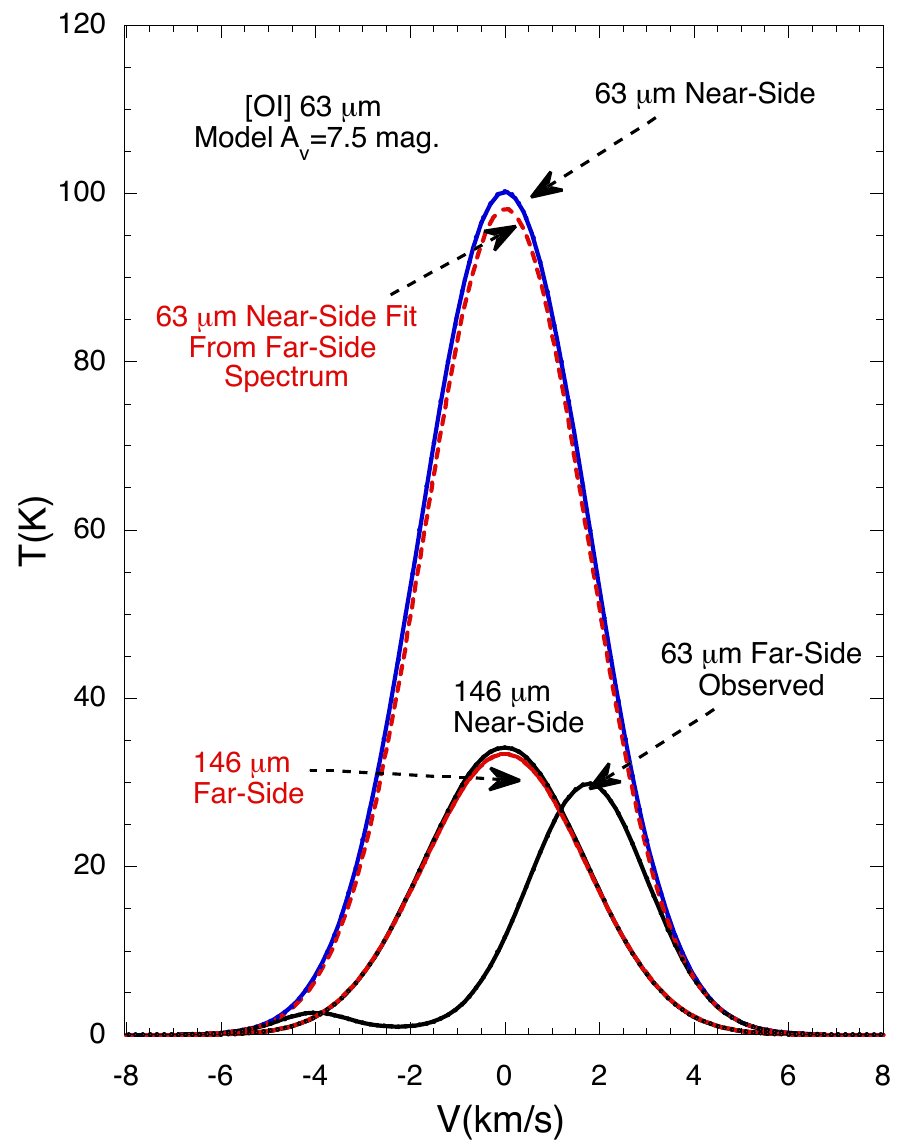}
\centering
\caption{Gaussian fits to a Meudon model of \OI\ in which the total extinction is $A_v$ = 7.5 mag. (see text). The near--side 63 \um\ spectrum is recovered based on maximizing the peak intensity while varying the velocity blocking window of the  far--side (observed) 63 \um\ spectrum.  The recovered spectrum differs by only $\simeq$3\% relative to the near--side configuration spectrum.  
\label{fig:FigA3}
}
\end{figure*}

The accuracy with which the unabsorbed emission can be recovered is quite good, with an underestimate of only 7\% in the peak antenna temperature in this example.  
The error is largely a result of the limited range of velocities for which data could be used given the very large central optical depth of the absorption produced in the low--excitation region, for the far--side cloud is between the emission region and the observer.   
The situation is even better for less--absorbed spectra, such as the ``outer' (-56,+55), (-29,+27), and (+28,-30)) positions observed, where the peak foreground optical depth is $\le$3.  
A model similar to that described above has the same heating source with total extinction of 7.5 mag.  
The line center velocities and line widths are the same, but the transitions are at 2.5 mag. and 4.0 mag. from the heated boundary.  
The total optical depth is less than half of that in the $A_v$ = 15 mag. model due to the greater fractional decrease in the size of the low--excitation region. 
The results for this case are shown in Figure \ref{fig:FigA3}.  
The peak antenna temperature of the Gaussian fitted to the wings of the far--side 63 \um\ line is less than that of the line in the near--side configuration by only 3\%.  

The Meudon PDR code produces noiseless spectra, and the Gaussian fits to the ideal Meudon  model 63 \um\ spectra are in consequence very good, additionally confirming that the emission line profiles are Gaussian, despite possible concerns about saturation broadening in the 63 \um\ line.
The fits to the actual data have additional errors due to the noise in the spectra. 
However, as seen in Table~\ref{tab:W3_all_gaussfits}, at the 1$\sigma$ level the errors in the fits to the observational data are, at worst, about 10\%.

\bibliography{bibdata}

\end{document}